\documentclass[apj]{emulateapj}
\usepackage{apjfonts}
\usepackage{amsmath}

\newcommand\healpix{HEALPix~}
\newcommand{\bm}[1]{\ensuremath{\mbox{\boldmath $#1$}}}

\shorttitle{Simulations of the Microwave Sky}
\shortauthors{Sehgal et al.}

\begin{document}

\title{Simulations of the Microwave Sky}

\author{Neelima Sehgal\altaffilmark{1}, Paul Bode\altaffilmark{2}, Sudeep Das\altaffilmark{2,3}, Carlos Hernandez-Monteagudo\altaffilmark{4}, Kevin Huffenberger\altaffilmark{5}, Yen-Ting Lin\altaffilmark{6}, Jeremiah P. Ostriker\altaffilmark{2}, and Hy Trac\altaffilmark{7}}

\altaffiltext{1}{Kavli Institute for Particle Astrophysics and Cosmology, Stanford, CA 94305}
\altaffiltext{2}{Department of Astrophysical Sciences, Princeton University, Princeton, NJ 08544}
\altaffiltext{3}{Department of Physics, Princeton University, Princeton, NJ 08540}
\altaffiltext{4}{Max Planck Institut fuer Astrophysik, 85741 Garching bei Muenchen, Germany}
\altaffiltext{5}{Department of Physics, University of Miami, Coral Gables, FL 33146}
\altaffiltext{6}{Institute for the Physics and Mathematics of the Universe, University of Tokyo, Japan}
\altaffiltext{7}{Harvard-Smithsonian Center for Astrophysics, Cambridge, MA 02138}

\begin{abstract}

We create realistic, full-sky, half-arcminute resolution simulations of the microwave sky matched to the most recent astrophysical observations.  The primary purpose of these simulations is to test the data reduction pipeline for the Atacama Cosmology Telescope (ACT) experiment; however, we have widened the frequency coverage beyond the ACT bands and utilized the easily accessible HEALPix map format to make these simulations applicable to other current and near future microwave background experiments.  Some of the novel features of these simulations are that the radio and infrared galaxy populations are correlated with the galaxy cluster and group populations, the primordial microwave background is lensed by the dark matter structure in the simulation via a ray-tracing code, the contribution to the thermal and kinetic Sunyaev-Zel'dovich (SZ) signals from galaxy clusters, groups, and the intergalactic medium has been included, and the gas prescription to model the SZ signals has been refined to match the most recent X-ray observations.  The cosmology adopted in these simulations is also consistent with the WMAP 5-year parameter measurements.   From these simulations we find a slope for the $Y_{200} - M_{200}$ relation that is only slightly steeper than self-similar, with an intrinsic scatter in the relation of $\sim 14\%$.  Regarding the contamination of cluster SZ flux by radio galaxies, we find for 148 GHz (90 GHz) only $3\%$ ($4\%$) of halos have their SZ decrements contaminated at a level of $20\%$ or more.  We find the contamination levels higher for infrared galaxies.  However, at 90 GHz, less than 20$\%$ of clusters with $M_{200} > 2.5 \times 10^{14} M_{\odot}$ and $z<1.2$ have their SZ decrements filled in at a level of $20\%$ or more. At 148 GHz, less than 20$\%$ of clusters with $M_{200} > 2.5 \times 10^{14} M_{\odot}$ and $z<0.8$ have their SZ decrements filled in at a level of 50$\%$ or larger.  Our models also suggest that a population of very high flux infrared galaxies, which are likely lensed sources, contribute most to the SZ contamination of very massive clusters at 90 and 148 GHz.  These simulations are publicly available and should serve as a useful tool for microwave surveys to cross-check SZ cluster detection, power spectrum, and cross-correlation analyses.
 
\end{abstract}


\def\bfof{b_{\rm FoF}}
\def\dtau{{\rm d}\tau}
\def\kpch{h^{-1}{\rm kpc}}
\def\Mfof{M_{\rm FoF}}
\def\Mpch{h^{-1}{\rm Mpc}}
\def\Msun{M_\odot}
\def\Msunh{h^{-1}M_\odot}
\def\Mvir{M_{\rm vir}}
\def\Omegab{\Omega_{\rm b}}
\def\Omegal{\Omega_\Lambda}
\def\Omegam{\Omega_{\rm m}}
\def\sigmaT{\sigma_{\rm T}}
\def\Rvir{R_{\rm vir}}
\def\Tvir{T_{\rm vir}}

\section{INTRODUCTION} \label{Neelima1}

Simulations of the microwave sky play a crucial role in accurately extracting constraints on cosmology potentially measurable by current and upcoming microwave background observatories, such as the Atacama Cosmology Telescope (ACT)\footnote{http://www.physics.princeton.edu/act/}, the South Pole Telescope (SPT)\footnote{http://pole.uchicago.edu/}, and the Planck satellite\footnote{http://www.sciops.esa.int/PLANCK/}.  These surveys will have complex data processing pipelines and analysis strategies, and any artifacts caused by any one of the many analysis steps can best be controlled by simultaneously running simulations of the microwave sky, with known cosmological and astrophysical inputs, through the same analysis pipeline.  These microwave background experiments, which will have finer resolution and higher sensitivity than achieved to date, can potentially constrain cosmological parameters via their measured microwave power spectrum, higher order correlation functions, Sunyaev-Zel'dovich (SZ) cluster detections, and cross-correlation studies of various microwave components.  Moreover, these microwave data sets can be combined with surveys at other wavelengths to yield additional constraints.  As such, the accurate recovery of known cosmological inputs, through the variety of methods mentioned above, must be tested with simulations.  This necessitates high-resolution, cosmological-scale, microwave simulations that include the most accurate astrophysical components as our current observations will allow.

The microwave sky consists primarily of the following components: (1) the primordial microwave background, which is lensed by the intervening structure between the last scattering surface and observers today, (2) the thermal SZ signal from the hot gas in galaxy clusters, groups, and the intergalactic medium, (3) the kinetic SZ signal from the bulk motion of galaxy clusters, groups, and the intergalactic medium, (4) higher-order relativistic corrections to the thermal and kinetic SZ signals, (5) a population of dusty star forming galaxies that emit strongly at infrared wavelengths but have significant microwave emission, (6) a population of galaxies that emit strongly at radio wavelengths, including active galactic nuclei, which also have significant emission at microwave frequencies, and (7) dust, synchrotron, and free-free emission from the Galaxy.

The simulations discussed here include the above components, mapped at six different frequencies: 30, 90, 148, 219, 277, and 350 GHz.  The frequencies 148, 219, and 277 are the central ACT observing frequencies, as these simulations are used by the ACT team to inform their data analysis pipeline.  The additional frequencies have been added to widen the applicability of these simulations to other microwave background experiments.   The only components listed above that have been excluded from the microwave maps are the Galactic synchrotron and free-free emission.  These two components are not expected to be significant at ACT frequencies, and we leave their inclusion to future work.  The maps themselves are full-sky \healpix maps of resolution 0.4 arcminutes (Nside=8192) and consist of full-sky maps of each component separately as well as a combined sky map, at each frequency.  These \healpix maps can be interpolated into flat sky maps using the bilinear interpolation subroutine included in the HEALPix distribution \citep{Gorski2005}.  In addition, catalogs of all the halos in this simulation and their various properties, as well as catalogs of the individual infrared and radio galaxies that populate the halos, are provided.  These simulations are public and can be downloaded from http://lambda.gsfc.nasa.gov/toolbox/tb\_cmbsim\_ov.cfm.

There are of course many simulations of components of the microwave sky (e.g. \citealt{Gawiser1997,Metzler1998,Gawiser1998,Sokasian2001,White2002,daSilva2004,Diaferio2005,Motl2005,Nagai2006,Hallman2007,Pfrommer2007,Roncarelli2007,Schafer2007,Leach2008,Pace2008,Shaw2008,Wik2008,Carbone2009,Peel2009}).  The simulations discussed here include the components mentioned above, in an internally consistent manner, matched to the most recent observations.  These simulations have improved upon the previous simulations discussed in \citet{Sehgal07} in the following ways: \\

\begin{itemize}
\item{radio and infrared galaxy populations are clustered and correlated with the galaxy cluster populations}
\item{the primordial microwave background has been lensed by the dark matter structure in the N-body simulation via a ray-tracing code}
\item{the contribution to the kinetic SZ signal from the intergalactic medium has been included, in addition to that from galaxy clusters}
\item{the gas prescription to model the SZ signals has been refined to match the most recent X-ray observations}
\item{maps have been created at a wider range of frequencies to increase the applicability to other microwave background experiments}
\item{standard \healpix format for the maps has been adopted}
\end{itemize}

We give a brief summary of the key elements of these simulations below, which we expand upon in greater detail throughout \S 2 and \S 3.  

The large-scale structure simulation was carried out using a tree-particle-mesh code \citep{bode.ostriker.ea:2000, bode.ostriker:2003}, with a simulation volume of $1000\ \Mpch$ on a side containing $1024^{3}$ particles.   The cosmology adopted is ($\Omega_b, \Omega_m, \Omega_{\Lambda}, h, n_s, \sigma_8$) = (0.044, 0.264, 0.736, 0.71, 0.96, 0.80), consistent with the WMAP 5-year results \citep{Komatsu2009}.  The mass distribution covering one octant of the full sky was saved, and halos with a friends-of-friends mass above $1 \times 10^{13} M_{\odot}$ and with a redshift below $z=3$ are identified. The octant is then replicated to create a full-sky simulation.  Thus the large-scale structure components in these simulations are unique for one octant, while the primary microwave background and Galactic dust are modeled for the full sky.   

The primary microwave background has the same cosmology as the large-scale structure and matches the WMAP 5-year Internal Linear Combination map \citep{Bennett2003b, Hinshaw2009} on large-scales ($l<20$).  The microwave background is then lensed by the large-scale structure in the simulation with a ray-tracing code \citep{das.bode:2008}.  

The Sunyaev-Zel'dovich signal is derived by adding a gas prescription to the N-body halos which assumes a polytropic equation of state and hydrostatic equilibrium.  This model, which is described in more detail in \citet{BodeOV2009}, adjusts four free parameters which are calibrated against X-ray gas fractions as a function of temperature from the sample of \citet{Sun2009} and \citet{Vikhlinin06}.  These four free parameters are the star formation rate (which depends on the halo mass and redshift), non-thermal pressure support (from cosmic rays or magnetic fields), dynamical energy transfer from dark matter to gas (a depletion factor which lowers the average baryon fraction to less than the cosmic mean), and feedback from active galactic nuclei (which is tied to the star formation rate).  This model describes the gas prescription for halos with friends-of-friends mass $> 3 \times 10^{13} M_{\odot}$ and $z<3$.

Smaller halos down to $1 \times 10^{13} M_{\odot}$ and the intergalactic medium for $z<3$ are treated with a different method.
The velocity dispersion of a given simulation particle is calculated using its neighbors, and this velocity dispersion is related to the pressure (and thus the thermal SZ signal) via a proportionality constant derived from hydrodynamical simulations \citep{Trac09}.  The kinetic SZ is calculated from the line-of-sight momentum of the particles.  For the $3<z<10$ Universe, the temperature is assumed to be isothermal at $10^{4} K$, and mass density and momentum shells at various redshift slices are used to construct the thermal and kinetic SZ signals. 

The infrared source model has six parameters used to populate halos, and is partially based on \citet{righi}.  This model satisfies the following observational constraints:  the COBE/FIRAS background as in \citet{firas} is an upper limit on the total infrared intensity, the effective spectral index for the spectral intensity is $\sim 2.6$ between 145 and 350 GHz  \citep{knox04, dunne}, source counts are compatible with SCUBA counts at 353 GHz \citep{coppin}, source counts and clustering are compatible with BLAST measured counts and power spectrum \citep{blast_counts,Viero09}, and the power of the infrared sources is below the point source upper limit derived by ACBAR at 150 GHz, fixing $\sigma_8$ to the WMAP 5-year measurement \citep{reichardt09}.

The model for the radio sources is made by creating a halo occupation distribution and luminosity distribution for radio sources at 151 MHz to match the 151 MHz radio luminosity function (RLF) at $z\sim 0.1$.  151 MHz was chosen because at such low frequencies the RLF is dominated by the steep spectrum lobes of the radio sources (as opposed to the flatter spectrum cores) and is thus less sensitive to biases due to orientation effects.  Once the 151 MHz RLF is matched at low redshifts, the radio sources are divided into two populations, similar to the types I and II of the \citet{fanaroff74} classification. These are given differing number density evolutions with redshift, and the RLF at higher redshift is matched to observations.  We use a relativistic beaming model to separate how much of a source's flux at 151 MHz is from the core as opposed to the lobes, and we adjust the shape of the radio core spectral energy distribution to match source counts at $\nu \gg 151$ MHz, where the core dominates \citep{Lin2009}.

The Galactic dust emission maps are from the ``model 8'' prediction from \citet{Finkbeiner1999}, which \citet{Bennett2003b} show to be a reasonable template for dust emission in the WMAP maps. 

In \S \ref{sec:sims}, we discuss the construction of each component of these simulations and comparison to observations in more detail.  In \S \ref{sec:discuss}, we discuss further properties of the simulated maps, and in \S \ref{sec:concl}, we summarize and conclude.

\section{SIMULATIONS}\label{sec:sims}

\subsection{Primary Unlensed CMB} \label{sec:}

We follow the method of \citet{Sehgal07}, constructing a primary microwave background map with the observed large-scale structure and small scale structure consistent with theoretical expectations.  For large
scales, we use the WMAP 5-year Internal Linear Combination (ILC) map \citep{Bennett2003b, Hinshaw2009}.  For $l<20$, we take the harmonic coefficients ($a_{lm}$) from the ILC map with no modification. At smaller scales, where the ILC map is smoothed significantly, a Gaussian random realization is added, so that the ensemble average power equals the theoretical power spectrum which maximizes the WMAP 5-year likelihood. Harmonic coefficients are computed to $l_{\rm max} = 8192$, and we synthesize a map with HEALPix parameter $N_{\rm side} = 4096$ (0.859\arcmin~pixels).  Then we convert to a map with HEALPix parameter $N_{\rm side} = 8192$ (0.429\arcmin~pixels) using the \texttt{\texttt{alter\_alm}} subroutine included in the HEALPix distribution \citep{Gorski2005}.  The lower resolution ($N_{\rm side} = 4096$) map is used to generate the lensed microwave background map (see \S~\ref{sec:lensing}), as that is more efficient than working with the $N_{\rm side} = 8192$ map directly and results in no significant loss of accuracy.  The lensed map is then also converted to an $N_{\rm side} = 8192$  map.

\subsection{Large-Scale Structure Simulation}\label{sec:lsssim}

A simulation of the large-scale structure of the Universe is
performed using the same methods as previously described in
\citet{bode.ostriker.ea:2007} and \citet{Sehgal07}, and we refer
the reader to their papers for additional details. Within a periodic
box of comoving side length $L = 1000\ \Mpch$, $1024^3$ dark matter
particles are evolved using the Tree-Particle-Mesh (TPM) N-body code
\citep{bode.ostriker:2003}. The particle mass is $6.82\times10^{10}\
\Msunh$ and the gravitational spline softening length is set to
$\epsilon = 16.276\ \kpch$. This simulation has been previously used by
\citet{BodeOV2009} in developing a model for the intracluster medium.

A putative observer at redshift zero is placed at one corner of the
periodic box, and the mass distribution along a past light cone,
covering one octant (box coordinates $x,y,z > 0$) of the sky, is
saved as the simulation is running.  To create full-sky maps
this octant is replicated, as described in \S \ref{sec:discuss}.

At each simulation time step, spanning the redshift range $z_1 > z > z_2$,
particles are found in a thin spherical shell at the comoving
distance $d$ corresponding to the light travel time to this observer,
spanning the range $d(z_1) > d > d(z_2)$.
For $z<3$ all of these particles are saved to disk and
for $z<10$ pixelized versions of the shells are saved, as described below.
There are 579 such shells in all;
the range $\Delta z = z_1 - z_2$ of each shell depends on the time
step size set by the TPM code.  It decreases from $\Delta z \approx 0.08$
at $z\approx 9$ to $\Delta z \approx 0.03$ at $z\approx 3$,
and to $\Delta z < 0.005$ for $z<0.8$.
For comoving distances larger than the box size, $d>1000\ \Mpch$,
there can be some duplication in structures as the periodic box
is repeatedly tiled.  This tiling is done without random
rotations to preserve the periodicity and prevent discontinuities in the
large-scale structure.
However, any repetition is usually at a different
redshift and evolutionary state. In cases where a halo
or filament appears twice at similar redshifts, it is
viewed at two different angles, making each projected image unique.

For all the shells ($z<10$)
particles are subdivided by angular coordinates using the HEALPix
scheme for pixelation of a sphere.
The projected mass and momentum in each pixel of the
shell are computed and saved to disk \citep[as described previously
in][]{das.bode:2008}.  These pixelized shells are used
to model the gravitational lensing (\S \ref{sec:lensing})
and high-redshift SZ distortion (\S \ref{sec:SZ}) of the microwave background.

For lower redshifts $z<3$, the positions and velocities of all particles
in the light cone are also saved to disk for postprocessing. The
approximately 55 billion particles thus saved occupy 1.3 TB of
disk space and are used to model the SZ effect (\S \ref{sec:SZ})
in detail. A friends-of-friends (FoF) halo finder, with a standard
linking length $\bfof$ equal to 1/5 of the mean interparticle spacing,
is used to identify dark matter halos and tag the particles belonging
to them. This halo catalog is used to model infrared (\S \ref{sec:ir})
and radio (\S \ref{sec:radio}) point sources.

\begin{figure}
\epsscale{1.2}
\plotone{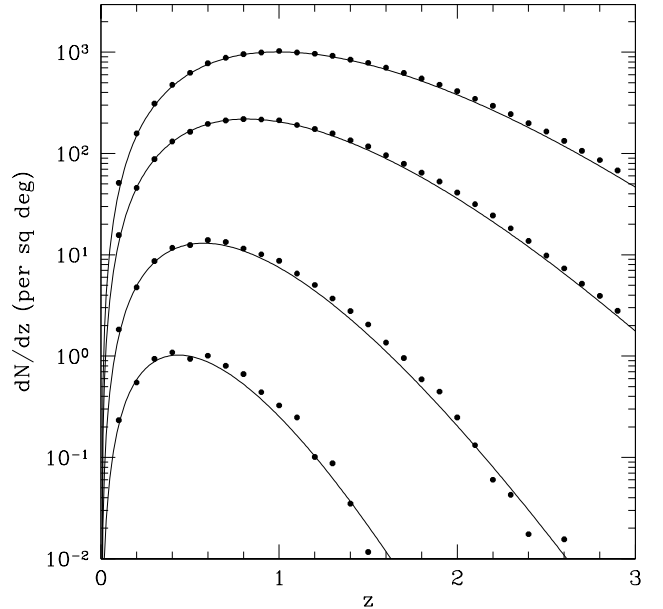}
\caption{The halo abundance above a minimum mass per unit redshift per
square degree measured from
the large-scale structure simulation (points) is compared with the semi-analytic
fitting formula (lines) from \citet{Jenkins2001}.  From top to bottom, the four sets of points are for the minimum masses $M_{\rm min}= 6.82\times10^{12}$, $2\times10^{13}$, $1\times10^{14}$, and $3\times10^{14}\ \Msunh$.}  \label{fig:dNdz}
\end{figure}

As shown in Fig.~\ref{fig:dNdz}, the halo mass functions measured from
particles in the light cone agree well with the semi-analytic fitting
formula from \citet{Jenkins2001} for a linking length $\bfof=0.2$.  The
number of halos above a minimum mass per unit redshift per square degree
is plotted for four minimum masses: $M_{\rm min}= 6.82\times10^{12}$,
$2\times10^{13}$, $1\times10^{14}$, and $3\times10^{14}\ \Msunh$. The
first limit, the equivalent of 100 particles, is the lowest mass of halos
that are populated with point sources. The second value corresponds to
massive halos with enough resolution to model the intrahalo gas in greater
detail. For any cluster survey, the minimum detectable mass is likely to
be a function of redshift, and the last two minimum mass values reflect
the range expected for upcoming SZ surveys. The data points in Fig.~\ref{fig:dNdz} are measured
using all halos in the octant to minimize sample variance, but then are
normalized per square degree.

\subsection{Lensing of the Primary CMB} \label{sec:lensing}

Gravitational lensing is a secondary anisotropy of the microwave background caused by the deflection of primary microwave background photons by  
intervening large-scale-structure potentials \citep[see][for a recent review]{lewis.challinor:2006}.  Lensing produces non-Gaussianities, 
smooths out features in the power spectrum and moves power from large to small scales. High resolution microwave background experiments, 
such as ACT, SPT, and Planck, will probe the scales corresponding to the multipole range  $500\lesssim \ell \lesssim 10000$ with unprecedented 
precision.  It will be necessary to include the effect of lensing for performing precision cosmological tests with such  datasets.  By studying the lensing distortions in the microwave sky, we should also be able to reconstruct the projected matter distribution or the deflection field  in the Universe \citep{hu.okamoto:2002,okamoto.hu:2003, hirata.seljak:2003*1, yoo.zaldarriaga:2008}. Such reconstruction will help break parameter degeneracies in the microwave background, making it sensitive to dark energy dynamics and neutrino mass \citep{smith.cooray.ea:2008,.zahn.ea:2009}. Also, the cross-correlation of microwave background lensing with various tracers of large scale structure will lead to better understanding of galaxy environments, structure formation, geometry and gravity on large scales \citep{acquaviva.hajian.ea:2008,das.spergel:2009*1,vallinotto.das.ea:2009}.  A realistic simulation of microwave lensing internally consistent with (i.e. based on the same large-scale-structure simulation as)  the other simulated secondary components, such as the thermal and kinetic SZ effects and infrared and radio point sources, is essential for training the power spectrum and parameter estimation pipeline. Such simulations are also necessary ingredients in  developing realistic lens reconstruction algorithms \citep{amblard.vale.ea:2004, perotto.bobin.ea:2009}. Here we present the salient aspects of a full-sky microwave background lensing simulation at arcminute resolution. \par
A simulation of the lensed microwave background essentially consists of remapping the points of the primary microwave background sky according to deflection angles derived from a simulated convergence field. Several approaches have been proposed for implementing the remapping as well as for generating the deflection field \citep{lewis:2005,das.bode:2008,carbone.springel.ea:2007,carbone.baccigalupi.ea:2008,basak.prunet.ea:2008}. In this work, we  follow \cite{das.bode:2008} in all respects, except that we create a full-sky convergence map (instead of the polar cap geometry adopted in that paper) and lens a full-sky primary microwave background map. The details of the algorithm can be found in  \cite{das.bode:2008} -- here we delineate the main steps:

\begin{figure}
\epsscale{1.3}
\plotone{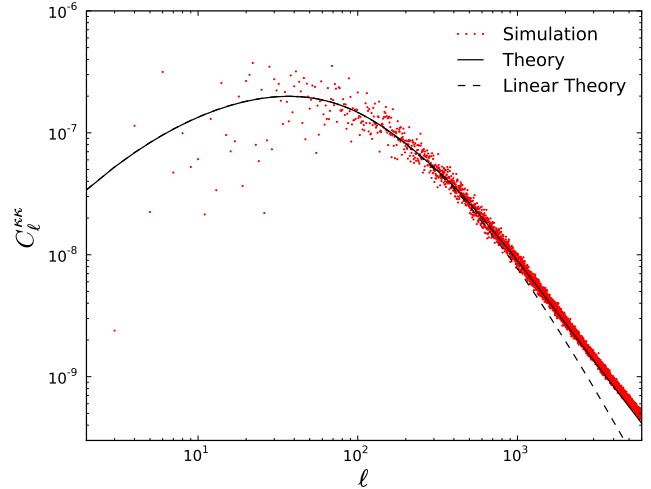}
\caption{Power spectrum of the simulated effective kappa map (dots) compared to theory, with (solid line) and without (dashed line) nonlinear corrections to the matter power spectrum. \label{sd_kappa_cls} }
\end{figure}

\begin{enumerate}

\item As described in \S\ref{sec:lsssim}, at each time step of the large-scale-structure simulation, the positions of the particles in a  thin shell (of width corresponding to the time step) are recorded onto the octant of a \healpix  pixelated sphere. Knowing the particle mass and the number of particles falling  into each pixel, we generate a surface mass density plane (expressed in mass per steradian) by dividing the total  mass  in a pixel by the solid angle subtended by it. We  then subtract the cosmological mean of the surface density to generate a surface overdensity on each shell.
\item We then generate the effective convergence map ($\kappa$) up to the maximum redshift in the large-scale-structure simulation ($z=10$) by combining the surface overdensities in the octant shells  with proper geometrical weighting  \citep[see, eqns. 15 and 20 in][]{ das.bode:2008}.   To accurately lens the microwave background, we need to include the contribution to the convergence from redshifts beyond the farthest simulated shell ($z>10$). We do this by adding in a Gaussian random realization for this extra convergence based on a theoretical convergence power spectrum (calculating the contribution to the convergence power from $z>10$ structure). At the end of this process, we have a realization of the effective convergence $\kappa(\hat n )$ on the octant. 
\item We replicate the octant to create a full-sky convergence map.  The detailed mapping of the octant to the full-sky is described in \S \ref{sec:discuss}. 
\item We invert the relation, $\kappa = \frac12 \nabla^2 \phi$, between the convergence and  the effective lensing potential $\phi$ in harmonic space to obtain the harmonic components of $\phi$, and generate the deflection field, $\bm{\alpha}(\hat n)$, using the relation $\bm{\alpha} = -\nabla \phi$.

\item We use the deflection field to  find the source-plane positions, $\bm{\theta}_s$, corresponding to the observed positions, $\bm\theta$, of the rays (centers of pixels) according to: $\bm\theta_s =\bm \theta+\bm\alpha$.
\item  Finally, we sample the unlensed microwave background at the source-plane positions using an accurate interpolation technique to produce the lensed microwave background map.
\end{enumerate}

\begin{figure*}
\begin{center}
\includegraphics[scale=0.45]{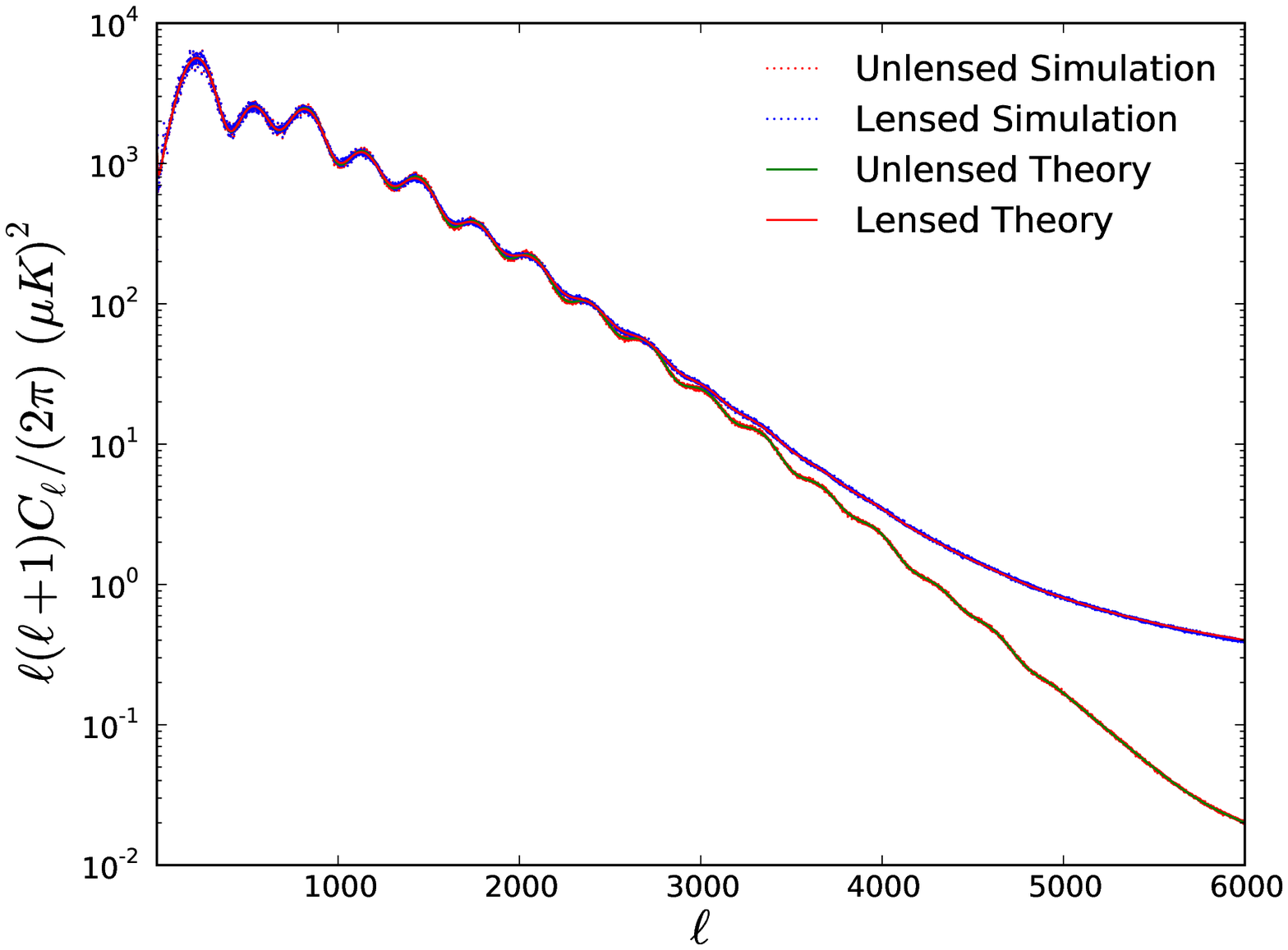}\includegraphics[scale=0.45]{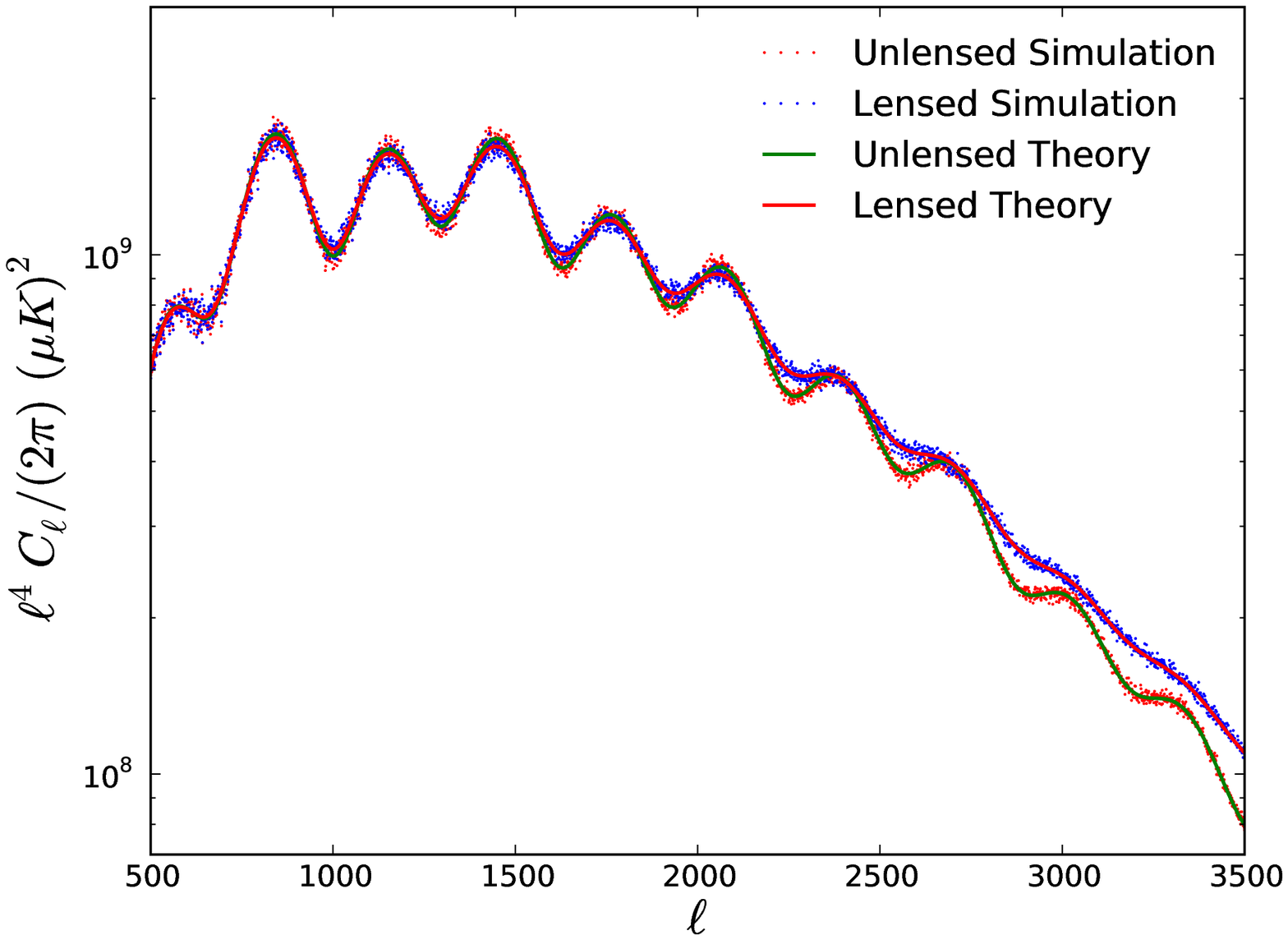}
\caption{\emph{Left}: Lensed and unlensed microwave background power spectra obtained from the simulation compared with the theoretical models.  The blue dots represent the power spectrum obtained from the simulated lensed microwave background map while the red solid line represents the theoretical lensed microwave background power spectrum from CAMB. The red dots represent the power spectrum of the input unlensed map and the green solid line is the theoretical input power spectrum. Note that the simulation and theory lines coincide to high precision. {\emph{Right:}} A smaller section of the spectra on the left. Note that the spectrum is multiplied by the fourth power of the multipole to make clearer the difference between lensed and unlensed spectra.}
\label{sd_lensedSpectra}
\end{center}
\end{figure*}

\begin{figure}
\begin{center}
\epsscale{1.35}
\plotone{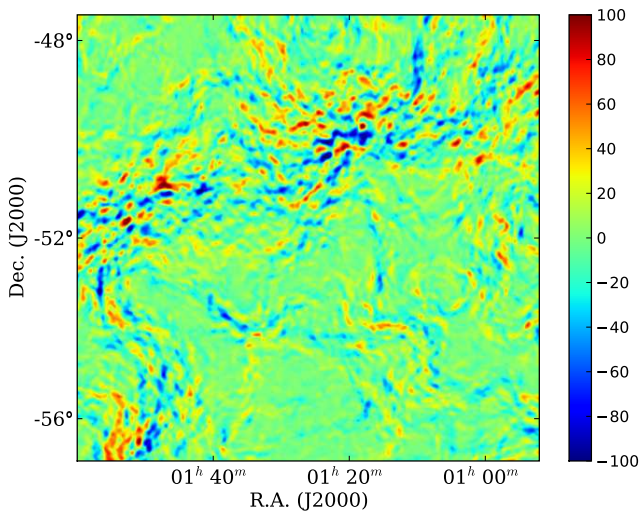}
\caption{Difference between the lensed and the unlensed primary microwave background map in a $\sim 10$ degree $\times~10$ degree patch. \label{sd_diff_map}}
\end{center}
\end{figure}

For  the results presented here, we have performed the above steps at the \healpix resolution $N_{\rm side} = 4096$.  Then we convert to a map with HEALPix parameter $N_{\rm side} = 8192$ using the \texttt{alter\_alm} subroutine included in the HEALPix distribution \citep{Gorski2005}.

Fig.~\ref{sd_kappa_cls}  compares the power spectrum of the simulated effective $\kappa$ map with two theoretical models: one  with and the other without nonlinear corrections  (following \citealt{smith.peacock.ea:2003}) to the matter power spectrum. The theoretical spectra have been computed with the publicly available CAMB\footnote{www.camb.info} code.  As expected, the simulated power spectrum agrees better with the theory curve with nonlinear corrections. There are two things worth noticing in the plot: the  large scatter at the low multipole end of the power spectrum, and the excess power in the simulation beyond $\ell \sim 2000$.  The scatter is expected because we generated the full-sky convergence map by replicating an octant.  The apparent high multipole excess is a consequence partly of the nonlinear corrections to the theory slightly underpredicting the nonlinear power, and partly of the shot noise due to the discrete nature of the particles.

We computed the power spectrum of the lensed microwave background map and compared it with theoretical expectations. Fig.~\ref{sd_lensedSpectra}  shows the  excellent agreement between  the theoretical power spectrum from CAMB and the power spectrum generated from the simulated lensed map.  For completeness, we also show the power spectrum of the unlensed input map and the theoretical power spectrum used to generate it. The simulation confirms the theoretical prediction that lensing smoothes out acoustic features in the microwave background and moves power from large to small scales.  Parenthetically, we would like to remark that the theoretical power spectra in Figs.~\ref{sd_kappa_cls}
 \& \ref{sd_lensedSpectra} were generated with CAMB using the parameter \texttt{k\_eta\_max\_scalar} set to  $20~\times $ \texttt{l\_max\_scalar}.  For  \texttt{k\_eta\_max\_scalar} = $2 \times$ \texttt{l\_max\_scalar} (which is normally used) the $\kappa$ power spectrum, as well as the lensed microwave background power spectrum, become inaccurate at high multipoles; the $\kappa$ power spectrum is suppressed beyond $\ell \sim 1000$, and the lensed microwave background power spectrum beyond $\ell \sim 4000$.
  
 A particularly interesting quantity is the difference between the lensed and the unlensed microwave background maps.  Fig.~\ref{sd_diff_map}  shows this quantity on a $7$ degree $\times\ 9$ degree patch projected onto a cylindrical-equal-area grid from the \healpix sphere.  The figure shows the striking large-scale non-Gaussian features that lensing induces on the microwave sky. Lensing reconstruction techniques use these large-angular-scale correlations of small-scale features in the microwave background to recover the lensing potential.

\subsection{SZ Signal} \label{sec:SZ}

The SZ effect is a secondary distortion of the microwave background that results when cosmic microwave radiation scatters against free electrons. It is commonly considered to have two main components \citep{SZ1970, SZ1972}. The thermal SZ (TSZ) term arises from inverse Compton scattering of the CMB with hot electrons, predominantly associated with shockheated gas in galaxy clusters and groups. The kinetic SZ (KSZ) term is a Doppler term coming from scattering with electrons having fast peculiar motions. Another common distinction is that the KSZ effect has a nonlinear component associated with the small-scale intracluster medium and a more linear component coming from the large-scale intergalactic medium (IGM). The signal arising from the latter was first calculated for the linear regime by \citet{OstrikerVishniac1986} and \citet{Vishniac1987} and is often referred to as the OV effect. Below, we first write down the formalism for the nonrelativistic limit and then the more general relativistic case.

Modeling the SZ effect requires knowing the number density $n_e$, temperature $T_e$, and velocity $v_e$ of the electron distribution. In the nonrelativistic limit, the change in the CMB temperature at frequency $\nu$ in the direction $\hat{n}$ on the sky is given by
\begin{equation}
\frac{\Delta T}{T_{\rm CMB}}(\hat{n})  = \left(\frac{\Delta T}{T_{\rm CMB}}\right)_{\rm tsz} + \left(\frac{\Delta T}{T_{\rm CMB}}\right)_{\rm ksz} = f_\nu y - b,
\label{eqn:SZnr}
\end{equation}
where the dimensionless Compton $y$ and Doppler $b$ parameters,
\begin{gather} \label{eqn:y}
y \equiv \frac{k_B \sigma_T}{m_e c^2}\int n_e T_e {\rm d}l =  \int \theta_{e}\dtau , \\ 
b \equiv \frac{\sigmaT}{c} \int n_e v_{\rm los}{\rm d}l =  \int \beta_{\rm los}\dtau ,
\end{gather}
are proportional to integrals of the electron pressure and momentum along the line of sight (los), respectively. The dimensionless temperature, line-of-sight peculiar velocity, and the optical depth through a path length ${\rm d}l$ are given by
\begin{gather}
\theta_e \equiv \frac{k_{\rm B}T_e}{m_e c^{2}} = 1.96\times10^{-3} \left(\frac{T_e}{{\rm keV}}\right), \\
\beta_{\rm los} \equiv \frac{v_{\rm los}}{c} = 3.34\times10^{-4} \left(\frac{v_{\rm los}}{100\ {\rm km/s}}\right), \\
\dtau \equiv \sigmaT n_e {\rm d}l = 2.05\times10^{-3} \left(\frac{n_e}{10^{-3}\ {\rm cm}^3}\right)\left(\frac{{\rm d}l}{{\rm Mpc}}\right),
\end{gather}
respectively. For a typical cluster, the Compton $y$ parameter is expected to be approximately an order of magnitude larger than the Doppler $b$ parameter.

The nonrelativistic TSZ component has a frequency dependence as specified by the function,
\begin{equation}
f_\nu \equiv x\coth(x/2) - 4 , \quad x\equiv h\nu/(k_{\rm B}T_{\rm CMB}) ,
\end{equation}
which has a null at $\nu\approx218$ GHz. The distortion appears as a temperature decrement at lower frequencies and as an increment at higher frequencies relative to the null. The nonrelativistic KSZ component is independent of frequency, but the sign of the distortion depends on the sign of the line-of-sight velocity. We chose the convention where $v_{\rm los}>0$ if the electrons are moving away from the observer.

In the general, relativistic case, the change in the CMB temperature at frequency $\nu$ in the direction $\hat{n}$ is given by
\begin{align}
\frac{\Delta T}{T_{\rm CMB}}(\hat{n}) = \int
\biggl[ & \theta_{e} (Y_0 + \theta_e Y_1 + \theta_e^2 Y_2 + \theta_e^3 Y_3 + \theta_e^4 Y_4) \biggr . \nonumber\\
\biggl . & + \beta^2 \left[\frac{1}{3} Y_0 + \theta_e \left(\frac{5}{6} Y_0 + \frac{2}{3} Y_1\right) \right] \biggr . \nonumber\\
\biggl . & - \beta_{\rm los} (1 + \theta_e C_{1} + \theta_e^2 C_2) \biggr] {\rm d}\tau,
\label{eqn:SZ}
\end{align}
where the $Y$'s and $C$'s are known frequency-dependent coefficients \citep{Nozawa1998}. Note that Eq.\ \ref{eqn:SZnr} contains only the first-order terms in Eq.\ \ref{eqn:SZ}, and that $f_\nu = Y_0$. We use Eq.\ \ref{eqn:SZ} to calculate the full SZ signal in these simulations. 

Sky maps of the SZ effect are made by tracing through the simulated electron distribution and projecting the accumulated temperature fluctuations onto a HEALPix grid with Nside = 8192. Using the saved simulation data described in \S \ref{sec:lsssim}, SZ maps are constructed with contributions from 3 components: massive halos at $z < 3$, low-mass halos and the intergalactic medium at $z < 3$, and the high-redshift Universe for $3 \leq z < 10$.

\begin{figure}
\epsscale{1.2}
\plotone{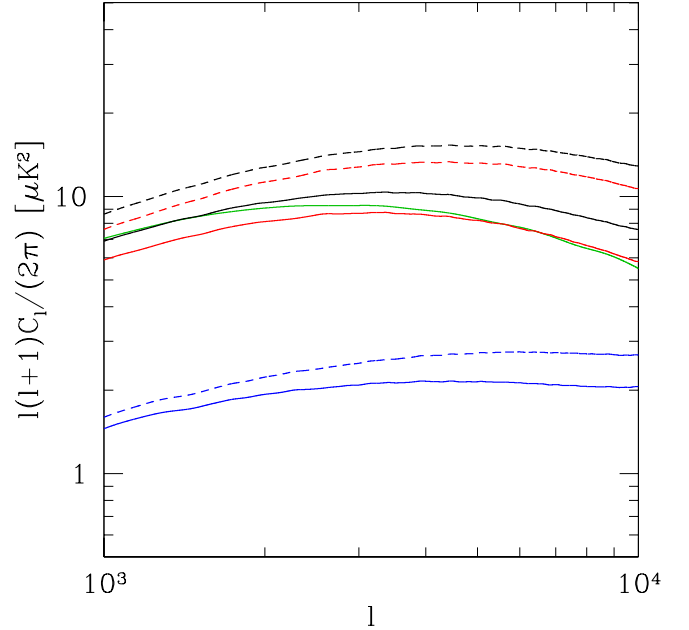}
\caption{Power spectra of the Sunyaev-Zel'dovich signals at 148 GHz.  Shown are the power spectra of the thermal SZ component (red), the kinetic SZ component (blue), and the full SZ signal (black).  The solid lines represent the model in this simulation, and the dashed lines represent a model with no feedback or star formation (adiabatic model \citealt{BodeOV2009,Trac09}).  The green line shows the comparison with the thermal SZ power spectrum derived analytically from \citet{Komatsu2002} and taken from the WMAP LAMBDA website (http://lambda.gsfc.nasa.gov/product/map/current/). }\label{szpower.eps}
\end{figure}

\subsubsection{Gas model for massive halos at $z<3$}
\label{sec:massivehalo}

The hot gas distribution associated with massive dark matter halos is calculated using a hydrostatic equilibrium model \citep{OstrikerBB2005, bode.ostriker.ea:2007, BodeOV2009}. Here we summarize the general method and refer the reader to \citet{BodeOV2009} for additional details, in particular on calibrating star formation and feedback against recent optical and X-ray observations. In the large-scale-structure simulation, massive dark matter halos with $\Mfof > 2\times10^{13}\ \Msunh$ and $z<3$ are postprocessed to include intrahalo gas and stars. These halos are expected to be the dominant contribution to the TSZ terms in Eqs.\ \ref{eqn:SZnr} and \ref{eqn:SZ} for the scales of interest ($1000\lesssim l \lesssim 10000$). Lower mass halos are accounted for differently, as we explain later.

For each massive halo, the collection of N-body particles tagged by the FoF finder are enclosed in a nonperiodic, Cartesian mesh that is centered on the center of mass, has one axis parallel to the line of sight, and a cell-spacing that is twice the N-body spline softening length. The particles are assigned using a cloud-in-cell scheme to define the matter density $\rho_{m}$ on the mesh. The gravitational potential $\phi$ is then calculated using a nonperiodic Fast Fourier Transform (FFT), and the location of the potential minimum is chosen as the halo center. This procedure preserves the concentration, substructure, and triaxiality of the dark matter halo.

Assuming hydrostatic equilibrium and a polytropic equation of state, the corresponding gas density $\rho_g$ and pressure $P_g$ on the same mesh are given by
\begin{gather}
\rho_g = \rho_{g,0}\psi^{1/(\Gamma-1)}, \\
P_g = P_{g,0}\psi^{\Gamma/(\Gamma-1)},
\end{gather}
respectively, where the dimensionless temperature variable,
\begin{equation}
\psi \equiv 1-\left(\frac{\Gamma-1}{\Gamma}\right)\left(\frac{\rho_{g,0}}{P_{g,0}}\right)(\phi-\phi_0),
\end{equation}
is related to the gravitational potential. \citet{BodeOV2009} adopted a polytropic index $\Gamma=1.2$ (consistent with the value of  $\Gamma$ adopted by other workers and the value found in hydrodynamic simulations) and determined the other two free parameters for the polytropic model, namely the central gas density $\rho_{g,0}$ and pressure $P_{g,0}$, by matching the gas mass fraction and temperature to recent X-ray observations \citep{Vikhlinin06, Sun2009}. The electrons are  assumed to have the same distribution of temperatures and velocities as the fully ionized gas.

For calculating the KSZ terms in Eq.\ \ref{eqn:SZnr} and \ref{eqn:SZ}, the gas velocity on the mesh can be modeled in two ways. The velocity field can be taken from the dark matter particles; this approach allows for velocity substructure, although the dispersion would be inconsistent with the hydrostatic equilibrium assumption. Alternatively, the entire gas distribution can be assigned a single, mass-weighted average peculiar velocity; this is the approach we adopt for self-consistency. Comparing the two options, we underestimate the temperature variance $<\Delta T_{\rm ksz}^2>$ from massive halos alone by only $\sim4\%$ when ignoring velocity substructure. Correspondingly, the KSZ angular power spectrum from massive halos alone is also underestimated, but only by $<1\%$ at $l=1000$ and $\sim6\%$ at $l=10000$. 

When a given halo is added to the SZ maps, mass and momentum are conserved by requiring that the baryonic mass be equal to the cosmic baryon fraction $\Omegab / \Omegam$ times the halo mass $\Mfof$. This is accomplished by first sorting the mesh cells by their gravitational potential, from most negative to least negative. The mesh is then traversed in order and cells are tagged until the accumulated baryonic mass, in gas and stars, reaches the mass limit. The outer radius of the gas is usually beyond the cluster virial radius.  Remaining cells do not contribute any SZ signal. This procedure results in a triaxial halo gas distribution with a physically-motivated cutoff.

We find that the gas associated with these massive halos makes up approximately half of the TSZ flux, but it is the dominant contribution to the TSZ angular power spectrum. It accounts for $\sim90\%$ of the total TSZ power for the scales of interest ($1000\lesssim l \lesssim 10000$). On the contrary, this gas accounts for only $\sim15\%$ of the total KSZ power because of the relative importance of the IGM to this signal, as discussed in \S \ref{sec:lowzigm} and \S \ref{sec:highzigm}.

\begin{figure}
\epsscale{1.35}
\plotone{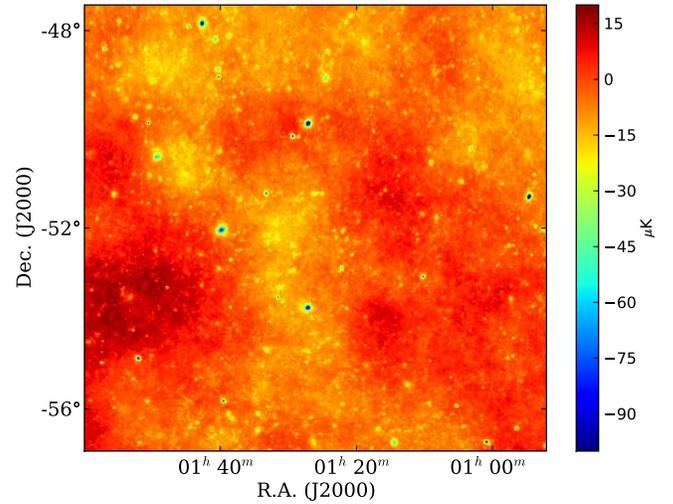}
\caption{The full Sunyaev-Zel'dovich signal at 148 GHz, including thermal and kinetic SZ contributions from both the galaxy clusters and the intergalactic medium, in a $\sim 10$ degree $\times~10$ degree patch. }\label{sz.ps}
\end{figure}

\subsubsection{Low-mass halos and intergalactic medium at $z<3$} \label{sec:lowzigm}

Low-mass halos and the intergalactic medium at $z<3$ are modeled using all other saved N-body particles not associated with the massive dark matter halos discussed in \S \ref{sec:massivehalo}. While we are able to identify halos down to $\Mfof = 6.82\times 10^{12}\ \Msunh$ (100 particles), there is insufficient resolution to accurately apply the hydrostatic equilibrium model. Instead, an alternative approach motivated by the virial theorem is adopted, one that can also be extended to particles not associated with identified halos. The intergalactic medium is expected to contribute only a small fraction of the TSZ signal, but it is very important to include it for calculating the KSZ signal.

According to the virial theorem, the internal energy of a virialized halo is twice its gravitational potential energy. During the formation of a halo, the dark matter undergoes violent relaxation and the gas gets shock-heated. The gas temperature can be estimated directly from the halo mass and used to model the TSZ signal \citep[e.g.][]{CohnWhite2009}, but we chose to calculate the gas temperature from the velocity dispersion of the dark matter. This approach is then applicable to all particles rather than just ones in identified halos.

For each particle, the local density $\rho_m$ and velocity dispersion $\sigma_v$ are calculated using 32 neighboring particles. For calculating the SZ terms, the effective pressure and momentum of the gas is taken as 
\begin{gather}
P_g = \frac{\Omegab}{\Omegam} \left[ \frac{(\gamma-1)\rho_m\sigma_v^2}{2\mu m_{\rm H}k_{\rm B}} \right], \\
(\rho v)_g = \frac{\Omegab}{\Omegam}(\rho v)_m , \label{eqn:gasmv}
\end{gather}
where $\gamma=5/3$ is the ideal gas adiabatic index and $\mu = 4/(3+5X_{\rm H}) = 0.588$ is the mean atomic weight for a fully ionized gas. Helium reionization is assumed to occur instantaneously at $z=3$. A temperature floor of $10^4$ K is imposed for the photoionized intergalactic medium.

When tracing through a given particle and projecting onto a HEALPix grid, the effective solid angle subtended by the particle is taken into account. A nearby, low-density particle can span a number of map pixels and simply assigning it to a single pixel will result in a large and incorrect temperature fluctuation. Mass and momentum are also conserved. Since the entire gas distribution at $z<3$ is accounted for, the maps have a smooth transition from the concentrated halos to the diffuse intergalactic medium. 

We find that the low-mass halos and the IGM make up approximately half of the TSZ flux, but contribute to the angular power spectrum only by $\sim10\%$ at $l=1000$ and  $\sim15\%$ at $l=10000$. This is in agreement with \citet{Hallman2007} who used a high-resolution hydrodynamic simulation and found that while roughly one-third of the TSZ flux comes from low-mass halos ($M< 5\times 10^{13}\Msun)$ and the IGM, these components add $\lesssim10\%$ to the TSZ angular power spectrum (for $\sigma_8=0.9$). \citet{Hallman2009} also found that the angular power spectrum of the filamentary warm-hot intergalactic medium \citep[WHIM][]{CenOstriker1999, Dave2001} is only $\lesssim1\%$ of the total.  Furthermore, \citet{CHM2006} found that while lower density gas ($\delta\lesssim10$) makes up approximately half of the baryon budget, it gives rise to only $\sim5\%$ of the TSZ flux in the local universe.

For low-mass halos in the mass range $6.82\times 10^{12} < \Mfof\ (\Msunh) < 2\times 10^{13}$, there are adequate particle numbers (100 to 282) to estimate the halo velocity dispersion. The TSZ signal from this component is sufficiently well approximated with the above method, considering that these halos contribute $\lesssim10\%$ to the TSZ angular power spectrum. However, for filaments with typical overdensities of 10, there are only 10 particles per comoving $(\Mpch)^3$. While it is difficult to calculate the velocity dispersion accurately, an approximate solution is acceptable since the filaments contribute only $\lesssim1\%$ to the angular power spectrum.

The KSZ signal from the low-mass halos and the IGM are relatively more important. We find that at $l=1000$, they account for $\sim15 \%$ of the total power, comparable to the contribution from massive halos, and at $l=10000$, they account for $\sim40\%$. The majority of the KSZ signal is from the high-redshift universe (3 < z < 10), which we discuss in \S \ref{sec:highzigm}. The KSZ signal is relatively less sensitive to resolution since the bulk peculiar momentum is well captured in the large-scale-structure simulation. While velocity substructure is not well resolved, we have already demonstrated in \S \ref{sec:massivehalo} that ignoring it only results in percent level differences. However, given the modest resolution of the large-scale-structure simulation, we caution against using the maps for detailed studies of the SZ signal from the WHIM and IGM.

\subsubsection{High-redshift Universe at $3\leq z<10$} \label{sec:highzigm}

The high-redshift universe at $3\leq z<10$ is modeled using the redshift shells containing the surface density fields for mass $\Sigma_m$ and line-of-sight momentum $\Sigma_{mv}$. At these higher redshifts, it is very reasonable to assume that the gas traces the matter distribution because the scales of interest are mostly still in the linear regime. Furthermore, since both the nonlinear mass and the collapsed mass fraction above a fixed mass rapidly decrease towards higher redshifts, the typical temperature of the gas is predominantly set by photoionization rather than shockheating.

For each redshift shell, the electron temperature, line-of-sight peculiar velocity, and column density are taken to be
\begin{gather}
T_e = 10^4\ {\rm K}, \\
v_{\rm los} = \frac{\Sigma_{mv}}{\Sigma_m}, \\
N_e = \left(\frac{\Omegab}{\Omegam}\right)\left(\frac{X_{\rm H}}{m_{\rm H}}+\frac{Y_{\rm He}}{m_{\rm He}}\right)\Sigma_m .
\end{gather}
Hydrogen reionization is assumed to occur instantaneously at $z=10$ and helium is only singly ionized. The Thomson optical depth for electron scattering is $\tau_{\rm T}\approx0.08$, consistent with the WMAP 5-year results \citep{Dunkley2009}. Since the electron distribution is warm rather than hot, the SZ temperature fluctuations are nonrelativistic and Eq.\ \ref{eqn:SZ} reduces to Eq.\ \ref{eqn:SZnr}. There is only a very small contribution to the TSZ signal from the warm electron distribution. The average Compton $y$ parameter is only $\sim10^{-7}$ over this redshift range. However, there is a substantial contribution to the KSZ signal because of the large, coherent velocities in the IGM.

Since the light cone is constructed by replicating and stacking from the periodic simulation box without random rotations, this procedure introduces excess power in the KSZ signal at $l\lesssim500$. The excess power shows up in the intergalactic medium component, but not in the halo component previously discussed in \S \ref{sec:massivehalo}. Therefore, the intergalactic medium component of the KSZ, from the full redshift range $0 < z < 10$, is filtered before adding it to the final SZ maps. A simple filter $w(l) = 1 - \exp[-(l/500)^2]$ is chosen to suppress the large-scale excess, but preserve power at $l>1000$. The remaining low-$l$ power is now more consistent with the level expected for the large-scale OV signal.

Since the simple filtering modifies the signal at $l<1000$, the maps should not be used to predict the KSZ signal at these scales. Furthermore, we have neglected to model the contribution from the inhomogeneous reionization epoch. This component is expected to be larger than the contribution from the reionized universe ($z\lesssim6$) at $l\lesssim1000$ \citep[e.g.][]{Santos2003, McQuinn2005, Zahn2005}.  The KSZ signal on these larger scales is subdominant compared to the lensed microwave background and to the TSZ signal at frequencies away from the null.

\subsection{Infrared Point Sources} \label{sec:ir}

Star formation in the Universe leads to the production of dust grains
which absorb the ultraviolet radiation in the environment and re-radiate it in the
infrared range. In the context of microwave background observations, the redshifted infrared
emission of these sources constitutes an important contaminant on small angular
scales at frequencies higher than $\sim$ 150 GHz. In this section, we
attempt to model the angular anisotropies that infrared (IR) galaxies
introduce in the microwave sky by combining the cosmological simulation presented above  
with a model for the infrared galaxy population, which is
partially based upon \citet{righi}.\\

\subsubsection{The basics of the infrared model}\label{sec:irBasics}

We first approximate the spectral energy distribution (SED) of the infrared
galaxies by a modified black body of the form
\begin{equation}
L_{\nu}\propto \phi(\nu, T_{dust}) \equiv \nu^{\beta}\left[B_{\nu}(T_{\mathrm{dust}})-B_{\nu}(T_{\mathrm{cmb}})\right],
\label{eq:lnu1}
\end{equation}
where $\beta$ is a spectral index, $T_{\mathrm{cmb}}$ is the primary microwave background monopole temperature, 
and  $T_{\mathrm{dust}}$ is the dust temperature evolving in redshift as in \citet{blain99}:
\begin{equation}
T_{\mathrm{dust}}(z)= \phantom{xx}
 \left[T_{\mathrm{0}}^{4+\beta}+T_{\mathrm{cmb}}^{4+\beta}\left((1+z)^{4+\beta}
  -1\right)\right]^{1/(4+\beta)}.
\label{eq:tdust1}
\end{equation}
Here $T_{\mathrm{0}}$ is the dust temperature if the galaxy was at $z=0$, and
the symbol $B_{\nu}(T)$ denotes the black body spectral intensity at temperature
$T$. As we shall see below, the free model parameters $\beta$ and
$T_{\mathrm{0}}$ will determine the spectral index for the effective power
law describing the change of the infrared intensity versus frequency,
$I_{\nu}\propto \nu^{\alpha}$ (i.e., $\alpha = \alpha(\beta,
T_{\mathrm{0}}) )$.

The star formation rate (SFR) in a given galaxy seems to be linearly
correlated with the dust luminosity \citep[Kennicutt relation,][]{kenni}.  At
the same time, there is observational evidence that the SFR reached a maximum
at epochs corresponding to $z\sim 2 - 4$ \citep{hopkinsb}. These two factors
combined suggest introducing a redshift window function for the infrared
luminosities in galaxies close to the measured SFR behavior. In our model, we
adopt the dependence 
\begin{equation}
W_{SFR}(z) \propto \exp{(-1/z)}\; \exp{\left(-\left(z/3.5\right)^2\right)}.
\label{equation}
\end{equation}
This window however gives more weight to the redsfhift range $z\in[1,2]$ if
compared to the data of \citet{hopkinsb}, due mainly to the absence of halos in
our simulation at $z>3$.

In practice, given the simulation halo catalog, our approach consists of populating each
halo with a number of galaxies with a given infrared luminosity. We adopt a mass
range limited by $M_1,M_2$ ($M_1 < M_2$), in such a way that halos below $M_1$
will host no infrared galaxies. Halos with masses $M \in [M_1,M_2]$ will typically
host {\em one} infrared galaxy of total infrared luminosity $L_{IR}(M,z)=
W_{SFR}(z) \cdot W_{cool}(M)\cdot L_{\star}\cdot(M/M_2)$. For more massive halos, the average
number of infrared galaxies will be given by $M/M_2$, each of them with a total infrared
luminosity given by $L_{IR}(M,z)=W_{SFR}(z) \cdot W_{cool}(M)\cdot L_{\star}$. The 
function $W_{cool}(M) = \exp(-M/M_{cool})$ accounts for the fact that bigger
halos need longer times to cool down before being able to form stars. $L_{\star}$ defines a luminosity parameter. The parameters
$M_1$, $M_2$, $L_{\star}$ and $M_{cool}$, combined with the spectral parameters
$\beta$ and $T_{\mathrm{dust}}$ introduced above, will define our infrared model. In the simulations,
the actual number of infrared galaxies present in a given halo is driven by a Poissonian realization 
of its average number. We also introduced an effective scatter of 15\% in the mean adopted value for 
the dust spectral index and in the infrared luminosity of each galaxy, partially motivated by the scatter in the effective spectral index of the intensity found in the analyses by \citet{knox04}.

\subsubsection{The angular power spectrum}

Below we calculate the angular power spectrum of our infrared model theoretically.  This allows us to approximate the power spectrum efficiently, after adjustments to the model parameters, without generating HEALPix maps.  
For each IR galaxy, the spectral luminosity $L_{\nu}$ can be computed from the
bolometric luminosity and the adopted form of the SED:
\begin{equation}
L_{\nu} = \frac{L_{IR}(M,z)}{\int d\nu\;\phi(\nu, T_{dust})} \phi(\nu, T_{dust}),
\label{eq:lnu2}
\end{equation}
where $T_{dust}$ is a function of redshift (Eq. \ref{eq:tdust1}). For a
flat Universe, the measured spectral flux is therefore given by 
\begin{equation}
F_{\nu} = \frac{L_{\nu'}}{4\pi\;r^2\;(1+z)},
\label{eq:lnu3}
\end{equation}
where $\nu = \nu'/(1+z)$ and $r$ is the comoving distance to the IR galaxy placed at redshift
$z(r)$. The {\em average} spectral intensity generated by the IR galaxy
population can then be written as the sum of the
contribution of all galaxies per unit solid angle:
\begin{align}
{\bar I}_{\nu} & = \int dr\; r^2\; \int dM\; \frac{dn}{dM}(r)\;N_g(M) \frac{a(r)L_{\nu'}(M,r)}{4\pi\;r^2} 
\label{eq:inuav}
\end{align}
The comoving halo mass function is given by $dn/dM$, and $N_g(M)$ provides the
average number of infrared galaxies in a halo of mass $M$.  The cosmological scale
factor is expressed by $a(r)$, and the typical spectral luminosity for each of the
infrared galaxies in the halo is given by $L_{\nu'}(M,r)$, with $\nu'$ being the
frequency observed at redshift $z(r)$.  $L_{\nu'}(M,r)$ is determined by the model described in \S \ref{sec:irBasics} and Eq. \ref{eq:lnu2}.

However, we are interested in the angular fluctuations of the infrared spectral
intensity, and this is linked to the angular variation of the number of infrared
sources.  The number of infrared galaxies is driven {\it a priori} by Poisson
statistics. However, halos tend to cluster in overdense regions of the large
scale matter density field, and this introduces another modulation in the infrared
galaxy number density that is known as the {\it correlation} term
\citep{zoltan00,song03,righi}. In terms of the spectral intensity $I_{\nu}$,
it is easy to show that the Poisson term for the angular anisotropies
generated by the infrared galaxy population can be written as \citep{righi}
\begin{equation}
C_l^P = \int dr\; r^2 \int dM\; \frac{dn}{dM}(r)\;N_g(M) \left(\frac{a(r)L_{\nu'}(M,r)}{4\pi r^2}\right)^2,
\label{eq:clp1}
\end{equation}
whereas the correlation or clustering term depends upon the halo power
spectrum:
\begin{equation}
C_l^C = \int \frac{dr}{r^2} P_m(k,r)
  \biggl[ \int dM \frac{dn}{dM}(r) b(M,r) N_g(M) \frac{a(r)L_{\nu'}(M,r)}{4\pi} \biggr]^2
\label{eq:clc1}
\end{equation}
where $k=(l+1/2)/r$. The bias factor $b(M,r)$ expresses how halos are distributed compared to dark
 matter, in such a way that $b^2( M,r)$ is defined as the ratio of the halo
 power spectrum and the matter power spectrum ($P_m(k,r)$). Here we have adopted the model by \citet{STbias}, which provides a good fit (few percent accuracy) to the simulated halo power spectrum. 
 The last two equations are obtained after using the Limber approximation, which should be
 accurate to better than 2\% for $l>20$ \citep[e.g.,][]{verde00}.
 
 \begin{figure*}
\begin{center}
\epsscale{1.1}       
\plotone{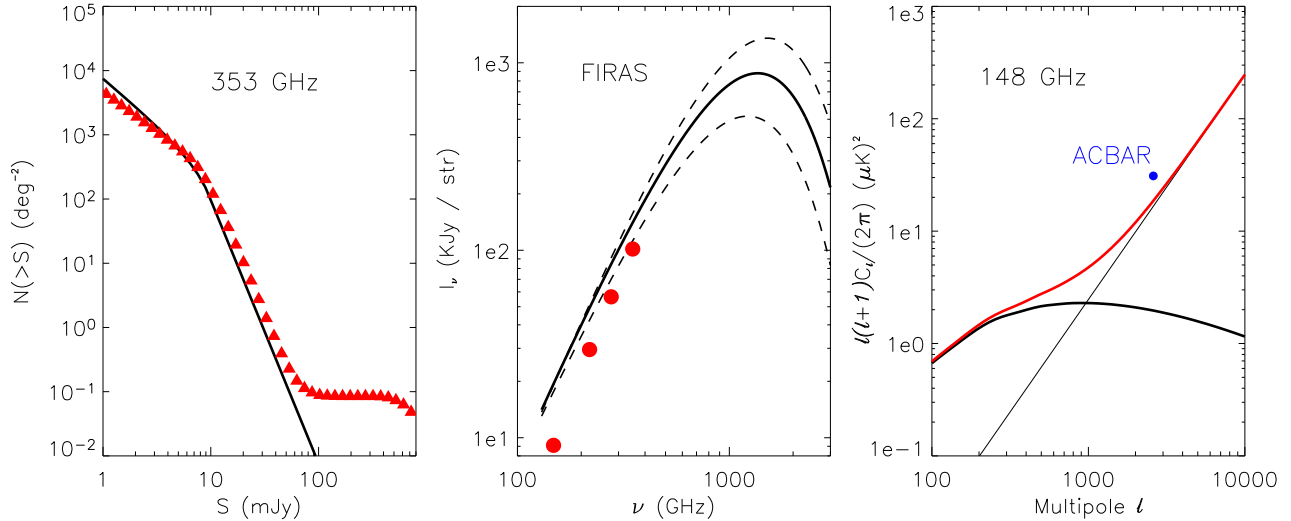}        
\caption[fig:]{{\it Left panel:} Cumulative source counts of our infrared model at 353 GHz versus flux (red triangles) . The solid line corresponds to the fit to the SCUBA data at the same frequency \citep{coppin}.  The high flux sources match the observed BLAST counts (\citealt{blast_counts} and BLAST team, private communication).  {\it Middle panel:} Contribution of our infrared source population to the total cosmic infrared background observed by COBE/FIRAS \citep{firas}. We show the three observing frequencies of ACT plus 353 GHz.  Dashed lines indicate $68\%$ confidence regions.  {\it Right panel:} Theoretical estimate of the angular power spectrum of our mock infrared source population. The thick black solid line corresponds to the clustering term, and the thin line to the Poisson term. The total is given by the red line, which is below the upper limit suggested by ACBAR at $l=2600$ \citep{reichardt09}.}
\label{fig:IR1}
\end{center}
\end{figure*}

\subsubsection{Observational constraints}

The parameter space is swept in such a way that our infrared model satisfies the
following observational constraints:
\begin{itemize}
\item The Cosmic Infrared Background (CIB), as estimated in \citet{firas} from
  COBE/FIRAS, is the upper limit for all total infrared intensity that our model
  provides.
\item The choice for spectral parameters $\beta$ and $T_{\mathrm{dust}}$ must
  be such that the effective spectral index for the spectral intensity ($I_{\nu}
  \propto \nu^{\alpha}$) is close to $\alpha \approx 2.6$ in the range $\nu
  \in [145, 350]$ GHz. This value is found in \citet{knox04}, and is based on the projection to high
  redshifts of the analyses of \citet{dunne} of a low redshift infrared galaxy
  sample. Nearby galaxies show harder spectral indexes, while high $z$ sources
  are closer to the peak of the dust SED and hence provide a shallower
  slope. The quoted value of the spectral index is practically
  equivalent to the slope of the CIB at the same frequency range. Thus,
  imposing $\alpha \approx 2.6$ makes the infrared background generated
  by our infrared sources have a slope close to that of the CIB in the frequency
  range $\nu \in [145, 350]$ GHz.
\item The source counts of our infrared model must be compatible to the source
  counts provided by SCUBA \citep{coppin} in the flux range $S_{353\;GHz} \in
  [1,50]$ mJy.  These source counts are in agreement with a number of other  submillimeter observations (e.g. \citealt{Austermann2009}).
\item The source counts of our infrared model must be compatible with the high flux
  infrared source population recently seen by BLAST (\citealt{blast_counts} and BLAST team, private communication). 
\item The clustered component of the the angular power spectrum should dominate over the Poisson term for $l<1000$ as observed by
 \emph{Spitzer} \citep{Lagache2007} and BLAST \citep{Viero09}.   
\item The angular power spectrum introduced by our IR source population should not exceed the upper limit
  derived from ACBAR observations at 150 GHz after fixing $\sigma_8$ to the WMAP 
  5-year best-fit value with a running spectral index ($\sigma_8 = 0.81$), i.e. $l^2C_l/(2\pi) <
  31 (\mu K)^2 $ at $l=2600$ \citep{reichardt09}.
\end{itemize}

 \begin{center}
\begin{deluxetable*}{ccccccc}
\tabletypesize{\scriptsize}
\tablecaption{Parameters for the two populations considered for the infrared point source model\label{tab:models}}
\tablehead{\colhead{Model} & \colhead{$\beta$} & \colhead{$T_{\mathrm{0}}$} & \colhead{$M_1$} & \colhead{$M_2$} & \colhead{$M_{cool}$} & \colhead{$L_{\star}$} \\
\colhead{}& \colhead{} & \colhead{($K$)} & \colhead{($M_{\odot}$)} & \colhead{($M_{\odot}$)} & \colhead{($M_{\odot}$)} & \colhead{($L_{\odot}$)} 
}
\startdata
Basic & $1.4$ & $25$ & $2.5\times 10^{11}$ & $3\times 10^{13}$ & $5\times 10^{14}$ & $1.25\times 10^{12}$\\
Very High Flux & $1.3$ & $40$ & $5\times 10^{14}$ & $1\times 10^{15}$& $8\times 10^{14}$ & $6\times 10^{14}$
\enddata
\end{deluxetable*}
\end{center}

\subsubsection{Final model}

We modeled two different source populations: the {\it basic} population essentially defines the properties of the whole IR model, whereas the {\it very high flux} population matches the high flux IR source counts detected by BLAST (BLAST team, private communication). Each of these two models requires a different choice for the parameter set, as shown in Table \ref{tab:models}. 

In order to avoid the limitation imposed by the minimum mass of the halo resolved in our simulation, we added a mock catalog of halos of masses between $10^9\; M_{\odot}$ and the minimum resolved halo mass, $M_{min}= 6.82 \times 10^{12}\; h^{-1}M_{\odot} = 9.61 \times 10^{12}\; M_{\odot}$. In order to do this, we assumed that the mass function in this mass range was dictated by the fit of \citet{Jenkins2001}. We used the population of low mass halos resolved in the simulation as tracers of the mock halos appended in this low mass range. We divided the mass range $[10^9\; M_{\odot}, M_{min}]$ in logarithmically spaced bins, and, in each of these bins, computed the average number of halos {\em per} resolved halo that is found in a thin mass interval centered at $M_c = 1.10 \times 10^{13}\; M_{\odot}$.  That is, for each resolved halo of mass close to $M_c$, we computed the average number of halos present in each of those low mass bins, and subsequently made a Poissonian realization of this average. The masses of these new mock halos are randomly distributed uniformly within the corresponding mass bin width. These mock halos were randomly distributed (with a Gaussian profile) in spheres of roughly 100 times the parent (resolved) halo virial radius, which should correspond to $\sim 10$ Mpc. Therefore, for each real halo of mass close to $M_c$, we populated a {\it cloud} of lower mass halos in an sphere of roughly 10 Mpc radius and centered on the parent halo itself. This procedure provides our simulation with a low mass halo catalog that is not Poisson distributed, but rather is correlated to the population of small but resolved halos close to $M=M_c$. In practice, only mock halos of mass above $M_1$ were generated, since they are the ones eventually hosting IR galaxies. 

The very high flux population consists of few ($\sim 300$ in the whole octant) infrared
galaxies above 200 mJy at 353 GHz. This very high flux population models the
tail observed by BLAST at 250, 350 and 500 $\mu$m  (BLAST team, private communication). 
The high flux tail should be at or above the 10 mJy level in the 145 -- 220 GHz channels, contributing significantly to 
the power spectrum. Some of this very high flux population includes sources believed to be lensed and magnified by the intervening structure \citep{Lima2009}. 

The total cumulative source counts above a given flux threshold for the sum of the basic and the very high flux populations are displayed in the left panel of Fig.~\ref{fig:IR1}. We show fluxes above 1 mJy (which is about SCUBA's detection threshold), although our  IR source population extends to dimmer fluxes. It is actually the numerous but dim IR source population that is responsible for most of the CIB. In our model, our IR source population generates between 50\% and 80\% of the total CIB detected by FIRAS (middle panel).  This is consistent with BLAST observations  \citep{Marsden2009, Devlin2009}, which indicate that most, and perhaps all, of the CIB is generated by infrared sources. 

In the right panel, we display the expected angular power spectrum of our total IR source population. At very high multipoles the Poisson term must dominate, whereas the clustering term takes over at $l \sim 1,000$.  Again due to the construction of the light cone by replicating and stacking the periodic N-body simulation box without random rotations, excess power is introduced in the infrared point source signal at $l\lesssim300$, as was the case for the kSZ signal from the intergalactic medium (see \S~\ref{sec:highzigm}).  We do not filter this signal to suppress the low-$l$ excess as we did for the kSZ, as this signal consists of discrete sources instead of a smooth component, and such a filter would disturb the real space point source distribution.  For $l>300$, the theoretical expectation of the angular power spectrum coincides with the actual power spectrum from the simulated infrared map.

The particular values for the model parameters for each of the two
populations are displayed in Table \ref{tab:models}. Note that the masses in this table refer to friends-of-friends halo masses.  
For the basic population, the values of the dust temperature and infrared luminosity are not far from
those found by \citet{blast_radio} when comparing BLAST infrared sources
to radio and mid-infrared counterparts in the Chandra Deep Field South. For the
few galaxies in the very high flux population,
characteristic masses and infrared luminosities must be increased by roughly an
order of magnitude, whereas the typical dust temperature increases from 25 to
40 $K$.

\subsection{Radio Point Sources} \label{sec:radio}

Our model for radio point sources follows in spirit the treatment of the subject by \citet[][hereafter W08]{wilman08}. 
Similar to W08,we separate the radio-loud active galactic nuclei (AGNs) into two populations, a high-power family that evolves strongly with redshift, and a low-power one with mild redshift evolution. These two populations can be roughly identified with the type II and I classification of radio sources by \citet{fanaroff74}, respectively (however, see discussions in \citealt{willott01}).
We also follow W08 and \citet{jackson99} to employ a relativistic beaming scheme to separate the contributions to the total flux from the compact core and the extended lobes of radio sources.

\begin{figure}[h]
\epsscale{1.2}
\plotone{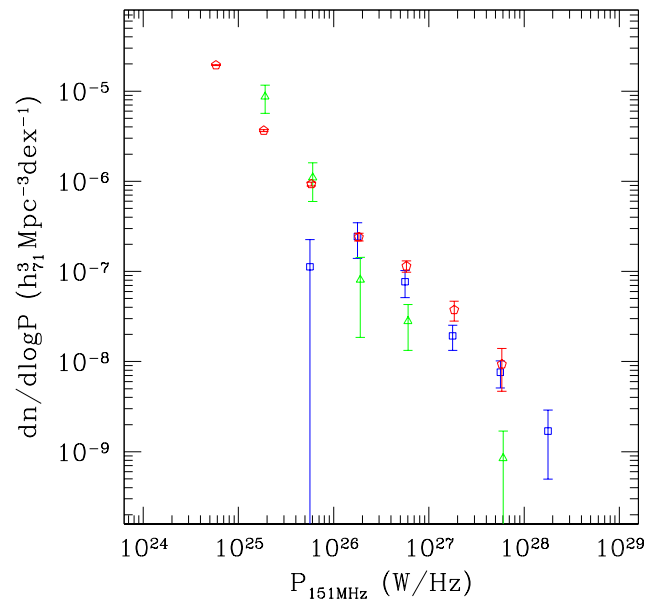}
\caption{
The 151 MHz RLF at $z\le 0.3$ constructed from the 3CRR survey, using data from \citet{laing83}. The blue squares are for FR IIs, and the green triangles are for FR Is. The red pentagons are the model prediction (sum of the two populations).}
\label{fig:rlf}
\end{figure}

Unlike W08, we do not consider radio-quiet AGNs and star-forming galaxies in the radio source model. As our primary goal is to simulate the microwave sky at relatively bright flux limits (e.g., $\gtrsim 0.1$ mJy), ignoring these populations has negligible effects. Furthermore, the infrared source modeling presented in \S\ref{sec:ir} is a self-contained model for the star-forming galaxies, and therefore the combined infrared and radio source models should adequately account for the majority of sources of interest to the current generation of microwave experiments.

Our model is a way to populate dark matter halos with radio sources in a Monte Carlo fashion, according to simple prescriptions on (1) the spectral energy distribution of the sources, (2) the number of radio sources as a function of halo mass and redshift, and (3) the (radio) luminosity distribution (LD) of the sources. Below we outline the most important aspects of the model.

\subsubsection{Compact versus extended emission of radio sources} 

The extended lobes of radio sources are characterized by a steep falling spectrum. On the other hand, the emission from the core region of a radio source is usually flatter and can exhibit a more complex spectral shape (see e.g., \citealt{lin09}). In addition, when the jet axis is close to the observer's line-of-sight, the fluxes from the core (and/or the base of the jets) will be Doppler-boosted.

Following W08, we assume the observed core-to-lobe flux ratio,  $R_{\rm obs}$, is related to the intrinsic value,  $R_{\rm int}$, via $R_{\rm obs} = R_{\rm int} B(\theta)$, where $\theta$ is the angle between the jet axis and the line-of-sight, $B(\theta) = [(1-\beta \cos \theta)^{-2} + (1+\beta \cos \theta)^{-2}]/2$, $\beta=\sqrt{1-\gamma^{-2}}$, and $\gamma$ is the Lorentz factor of the jet.

Given the  intrinsic luminosity $L_{\rm int} = L_{\rm c,int}+L_{\rm l,int}$ of a source, where the subscripts c and l refer to the core and lobes,
the beamed luminosity is then
\begin{align}
L_{\rm beam} & =  L_{\rm c,beam}+L_{\rm l,int}  \nonumber\\
& = (1+R_{\rm obs}) L_{\rm l,int} \nonumber\\
& = L_{\rm int} (1+R_{\rm obs})/(1+R_{\rm int})
\end{align}
We use empirically derived values of $R_{\rm int}$ and $\gamma$ from W08, and we assume a uniform distribution of $\cos \theta$ in order to separate $L_{\rm c,beam}$ and  $L_{\rm l,int}$ given $L_{\rm beam}$.

Since the compact core emission is usually observed to exhibit a complex spectral shape \citep{lin09}, we model it with a third-order polynomial, $\log f_\nu = \sum_i^{3} a_i (\log \nu)^i$, where $\nu$ is in GHz. We assume the extended sources obey $f_\nu \propto \nu^{-0.8}$, which is also consistent with observations \citep{lin09}.

\subsubsection{Halo occupation number and the 151 MHz RLF} 

Our first task is to reproduce the local 151 MHz radio luminosity function (RLF), which is effectively a convolution of the halo mass function, the halo occupation number, and the radio luminosity distribution.
The main reason for targeting the RLF at 151 MHz is because at such low frequencies, one can safely assume that the fluxes are mostly dominated by the steep spectrum, extended sources, and the resulting RLF is not very sensitive to biases due to orientation effects.  In addition, this RLF is not greatly affected by synchrotron/inverse Compton losses \citep{blundell99}.  For both FR I and FR II populations, we assume their halo occupation number (HON) is given by $N(M) = N_0 (M/M_0)^\alpha$. As for the LD, we assume it is independent of halo mass, and adopt two-segment power-laws that roughly mimic the observed RLF:\\
\begin{equation}
p(L) = \left\{ \begin{array}{ll}
                     (L/L_b)^m & \mbox{if $L>L_b$} \\
                     (L/L_b)^n  & \mbox{otherwise}
                     \end{array}
           \right.
\end{equation}

We explore the parameter space spanned by the HON and LD for both radio source populations, until the combined RLF fits the observed one at $z\sim 0.1$, which is based on the 3CRR survey (\citealt{laing83}).
A comparison of the model prediction and the observed 151 MHz RLF is shown in Fig.~\ref{fig:rlf}.

\begin{figure}[t]
\epsscale{1.2}
\plotone{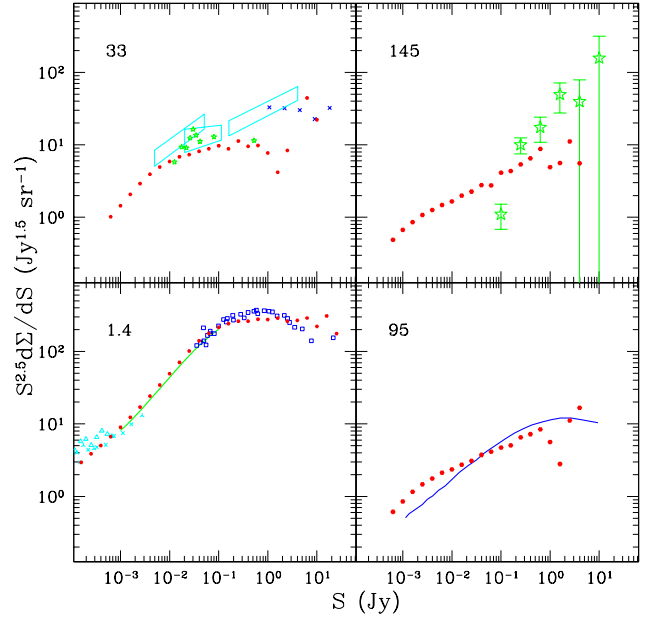}
\caption{Source counts at four frequencies (1.4, 33, 95, and 145 GHz). 
The model prediction (one realization of $a_i$'s - see Table \ref{tab:fr}) is shown as red solid points. All other symbols are observed source counts. The only exception is the curve in the 95 GHz panel, which is itself a model from \citet{dezotti05}, as shown in \citet{sadler08}. The data used in the 1.4 GHz panel are from \citet{katgert88,windhorst93,bondi03,huynh05}. The data for 33 GHz is from \citet{wright09,cleary05}. The data for 145 GHz is from ACBAR \citep{reichardt09}.}
\label{fig:rlnls}
\end{figure}

\subsubsection{Redshift evolution and source number counts} 

Once our model successfully reproduces the low redshift 151 MHz RLF, the radio sources are divided into two populations, roughly corresponding to the type I and II classification of \citet{fanaroff74}.  
These populations are given different number density evolutions with redshift, so that both the high-$z$ RLFs and source counts at 151 MHz can be fitted.  
For FR I sources, we consider an evolution of the form $(1+z)^\delta$ up to $z_p$ (and constant at $z>z_p$); for FR IIs, we find that a much stronger redshift evolution is needed and, for convenience, adopt an asymmetric Gaussian centered at $z_p$, with dispersions $\sigma_l$ and $\sigma_h$ at $z\le z_p$ and $z>z_p$, respectively.  (Note that there are many more FRIs and FRIIs at high redshift; for example, the comoving density of model FRII sources is $\sim 200$ times higher at $z_p=1.3$ compared to at $z\sim 0$.) In our model, $z_p$ is a free parameter, which is adjusted to match observations.

Having matched the high-$z$ 151 MHz RLF, we assume each source at 151 MHz is described by $L_{\rm beam} =  L_{\rm c,beam}+L_{\rm l,int}$, where the lobe component dominates.  Given $R_{\rm int}$ and $\gamma$ from W08, and $\cos \theta$ picked from a uniform distribution, we determine $L_{\rm c,beam}$.  We next adjust the shape of the core SEDs (by varying the coefficients $a_i$) so that the source counts at $\nu>151$ MHz are reproduced. These source counts at higher frequencies are dominated by the radio cores. In Fig. 9, we show a comparison of our model prediction (red solid points) with observed source counts at 1.4, 33, 95, and 145 GHz (see caption for  observation references). In general the model provides a good fit to the data\footnote{We have also compared our
source count predictions with observations at 0.15, 0.6, 2.7, 4.9, 8.4, 15, 20, 41, and 61 GHz, and found good agreement.}. We note that
for the 95 GHz panel, we are comparing our model to the predictions of the de Zotti et al. (2005) model.

\begin{deluxetable*}{ccccccccccccccccc}
\tablecaption{Model Parameters for Radio Sources}

\tablehead{
\colhead{Type} & \colhead{$R_{\rm int}$} & \colhead{$\gamma$} & \colhead{$a_0$} & \colhead{$a_1$\tablenotemark{$\dagger$}} & \colhead{$a_2$\tablenotemark{$\dagger$}} & \colhead{$a_3$\tablenotemark{$\dagger$}} & \colhead{$N_0$} & \colhead{$M_0$} & \colhead{$\alpha$} & \colhead{$L_b$} & \colhead{$m$} & \colhead{$n$} & \colhead{$\delta$} & \colhead{$z_p$} & \colhead{$\sigma_l$} & \colhead{$\sigma_h$} \\
\colhead{} & \colhead{} & \colhead{} & \colhead{} & \colhead{} & \colhead{} & \colhead{} & \colhead{} & \colhead{($h_{71}^{-1}M_\odot$)} & \colhead{} & \colhead{(W/Hz)} & \colhead{} & \colhead{} & \colhead{} & \colhead{} & \colhead{} & \colhead{} 
}

\startdata
FR I & $10^{-2.6}$ & 6 & 0 & $(-0.12,0.07)$ & $(-0.34,0.99)$ & $(-0.75,-0.23)$ & 1 & $4\times10^{13}$ & 0.1 & $10^{24}$ & $-1.55$ & $0$ & 3 & 0.8 & \nodata & \nodata\\
FR II & $10^{-2.8}$ & 8 & 0 & $(-0.12,0.07)$ & $(-0.34,0.99)$ & $(-0.75,-0.23)$ & 0.015 & $3\times10^{15}$ & 0.1 & $10^{27.5}$ & $-1.6$ & $-0.65$ & \nodata & 1.3 & 0.4 & 0.73
\enddata

\tablenotetext{$\dagger$}{We draw random uniform variates in this range when assigning the spectral shape for cores.}
\label{tab:fr}
\end{deluxetable*}

There are 24 free parameters in the model, which fit 24 observational constraints (12 RLFs at various frequencies and redshifts, and 12 source count observations at 12 frequencies).  (Note that $a_0$, $a_1$, $a_2$, $a_3$,  and $\alpha$ are set to be the same for both FRIs and FRIIs.)  Several of the parameters are degenerate; in Table~\ref{tab:fr} we present a parameter combination that can reproduce the results presented here.

For a more complete description of the model and further comparison with available observational constraints see \citet{Lin2009}.

\subsection{Galactic Emission} \label{sec:dust}

For the contribution of Galactic thermal dust emission, we use the
``model 8'' prediction from \citet{Finkbeiner1999}, an extrapolation
to microwave frequencies of dust maps from \citet{Schlegel1998}.  We use the software included with these maps to interpolate onto an oversampled HEALPix grid with   $N_{\rm side} = 8192$. 
 This model is a two-component fit to IRAS, DIRBE, and FIRAS data, and is shown by \citet{Bennett2003b} to be a reasonable template for dust emission in the WMAP maps. 

The Galactic synchrotron and free-free emission is comparable to the Galactic dust emission at $\sim 70$ GHz and becomes dominate/subdominant at lower/higher frequencies (see e.g. \citealt{Smoot1998}).   Thus these signals are not expected to be a significant contaminant for ACT or SPT (which surveys target regions away from the Galaxy), although they will be important to consider for the Planck low-frequency channels.  We leave the inclusion of these signals to later work, and note work by, for example, \citet{Leach2008}, which discusses the inclusion of these signals in microwave simulations for Planck.

\begin{figure}
\epsscale{1.35}
\plotone{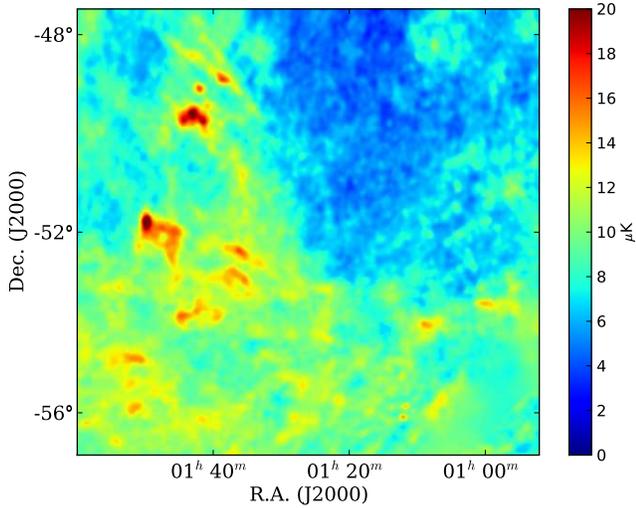}
\caption{The simulated Galactic dust emission in a $\sim 10$ degree $\times~10$ degree patch.}\label{dust.eps}
\end{figure}

\section{DISCUSSION} \label{sec:discuss}

\subsection{General Map Properties}

Below we list some general properties of the maps themselves.  The full-sky HEALPix maps for this simulation are in the Celestial coordinate system with RING ordering \citep{Gorski2005}.  The conversion between the HEALPix $\theta$ and $\phi$ coordinates and right ascension and declination is: $\phi = \rm{ra}, ~90-\theta = \rm{dec}$.  We store all maps in units of flux per solid angle (as Jy/sr).  To convert to temperature (as $\mu$K) we use the conversion:
\begin{equation}
\frac{dT}{dI}=\frac{c^2}{2 k_{B} \nu^{2}}\frac{(e^x - 1)^2}{x^2 e^x}  
\end{equation}
where $x=h \nu/k_{B} T_{CMB}$.  To convert from SI to more convenient units requires $\mu
$K sr/Jy = $10^{20}$ K sr m$^2$ Hz / W.  When calculating the power spectrum with Polspice\footnote{http://www.planck.fr/article141.html}  or anafast\footnote{http://healpix.jpl.nasa.gov/html/facilitiesnode6.htm}, a correction for the pixel window function is required.  

The properties of the halos that are specified in the halo catalog are redshift, ra, dec, position of the halo potential minimum, peculiar velocity, friends-of-friends mass, virial mass, virial gas mass, virial radius, integrated Compton $y$ parameter, integrated kinetic SZ, integrated full SZ at the six frequencies (the latter three integrated within the virial radius, $R_{200}$, and $R_{500}$), the stellar mass, and the central gas density, temperature, pressure, and potential.  In the radio and infrared galaxy catalogs, the properties given for each source are redshift, ra, dec, and the flux in mJy at the six frequencies.  

The large-scale structure octant is mirrored into full-sky maps as follows:

\begin{align}
\theta_{N} & = \theta_{0} \\ \nonumber
\phi_{N} & = \phi_{0} + n\frac{\pi}{2} \\  \nonumber
\theta_{S} & = \pi - \theta_{0} \\ \nonumber
\phi_{S} & = \frac{\pi}{2} - \phi_{0} + n\frac{\pi}{2} \\ \nonumber
\end{align}
where $n$ = 0, 1, 2, and 3, and the subscripts $N$ and $S$ refer to the northern and southern hemispheres of the map.  The subscript 0 refers to the coordinates in the original octant.  
For the power spectrum analysis, this method for repetition in the
Southern hemisphere is preferable to reflecting the North.  Although
simple mirroring eliminates real space discontinuities across the equator, it makes the simulated sky
an even function with respect to the polar axis.  In the spherical
harmonic transform, the associated Legendre functions are even or odd
depending on $l$.  Integrating mirrored hemispheres over them yields
zero for odd-$l$ harmonic coefficients.  Because rotating the sky
transforms harmonics with common $l$ into each other, odd-$l$ harmonics
will be zero regardless of the orientation of the mirroring.

\begin{figure}[h]
\epsscale{1.2}
\plotone{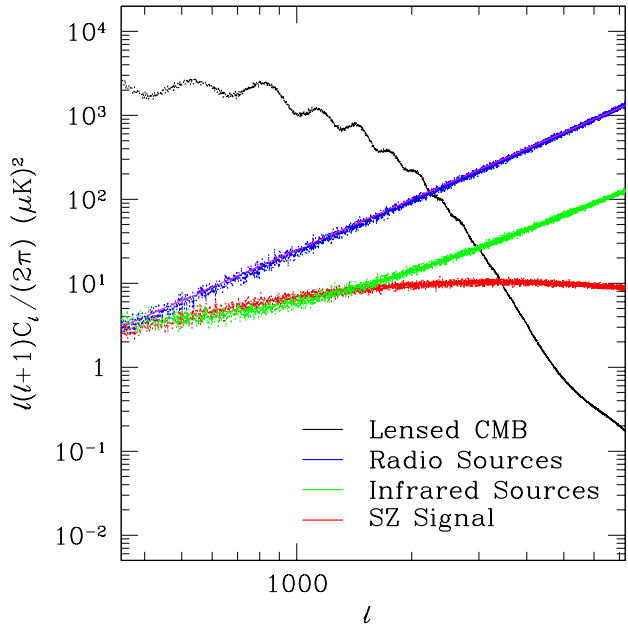}
\caption{Power spectra of the lensed microwave background (black), radio galaxies (blue), infrared galaxies (green), and full Sunyaev-Zel'dovich signal (red) in this simulation at 148 GHz.  The magenta line through the blue radio source power spectrum gives the radio source power from the \citet{Toffolatti1998} model scaled by 0.64, which was found by \citet{wright09} to be a good fit to the WMAP 5-year source counts.}
\label{fig:power}
\end{figure}

\subsection{Power Spectra of Simulated Components} 

In Fig.~\ref{fig:power}, we show the power spectra of the lensed microwave background, infrared galaxies, radio galaxies, and the full SZ signal from the simulated maps at 148 GHz.  The SZ power peaks at around $l \sim 3000$ and the clustering of the infrared galaxies is apparent for $l<1000$, consistent with measurements by \emph{Spitzer} \citep{Lagache2007} and BLAST \citep{Viero09}.  No infrared or radio sources have been removed in creating the power spectra, which were made with Polspice on the full-sky maps.  The power from the infrared and radio sources, without masking out bright sources, is comparable to the lensed microwave background between $l\sim2000$ and $l\sim3000$.  In Fig.~\ref{fig:power}, the magenta line through the blue radio source power spectrum gives the radio source power suggested by the \citet{Toffolatti1998} model scaled by 0.64, which was found by \citet{wright09} to be a good fit to the WMAP 5-year source counts.  The two power spectra are in excellent agreement.

\subsection{Y-M Relation and Scatter} 

We investigate the SZ flux versus cluster mass scaling relation for this simulation below.  An accurate calibration of this relation off observations is important for assessing the selection function of a potential SZ selected cluster sample via simulations.  The gas physics input into the simulations will determine the surface brightness of a given cluster based on its mass and redshift, which in turn determines how likely it is to be detected by a given observation strategy.  

To characterize the SZ flux - cluster mass scaling relation in this simulation we compare the integrated Compton $y$ parameter to cluster mass for the halos in the one unique octant.  The integrated Compton $y$ parameter is given by $Y=(\int y d\Omega) d_{A}^{2}(z)$, where $y$ is defined in Eq.\ \ref{eqn:y}, $d\Omega$ is the solid angle of the cluster, and $d_{A}^{2}(z)$ is the cluster's angular diameter distance.  
For the self-similar model, we expect a virialized halo of mass $M_{\mathrm{vir}}$ to have a virial temperature of 

\begin{eqnarray}
T_{\mathrm{vir}}\propto [M_{\mathrm{vir}} E(z)]^{2/3},
\end{eqnarray}
where for a flat $\Lambda$CDM cosmology
\begin{eqnarray}
E(z)=[\Omega_{m}(1+z)^3 + \Omega_{\Lambda}]^{1/2}.
\end{eqnarray}
If galaxy clusters were isothermal, then the Compton $y$ parameter would satisfy the relation 
\begin{eqnarray}
Y \propto f_{\mathrm{gas}} M_{\mathrm{halo}} T,
\end{eqnarray}
where $f_{\mathrm{gas}}$ is the cluster gas mass fraction.  Combining the above equations, we would expect 
\begin{eqnarray}
Y\propto f_{\mathrm{gas}} M_{\mathrm{vir}}^{5/3} E(z)^{2/3},
\label{eq:self-sim}
\end{eqnarray}
for the self-similar relation.  Deviations from the self-similar relation are not unexpected as clusters are not isothermal, not always in virial equilibrium, and $f_{\mathrm{gas}}$ need not be constant with variations in cluster mass and redshift.

\begin{figure}[h]
\epsscale{1.2}
\plotone{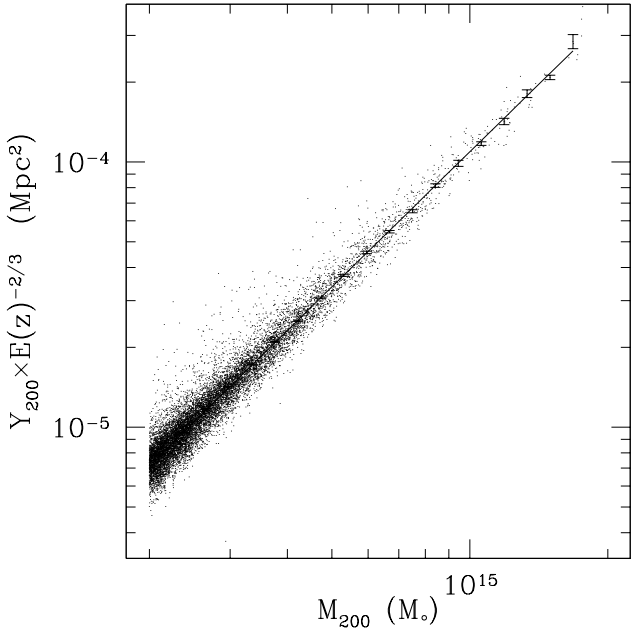}
\caption{$Y_{200}$ versus $M_{200}$ relation for all the halos from one octant of this simulation with $M_{200} > 2 \times 10^{14} M_{\odot}$.  The points with error bars represent bins of width $\Delta \rm{log} (M_{200}) = 0.05$ and errors on the bin mean.  The best-fit parameters for this scaling relation and the scatter are given in Table \ref{tab:ym}.}\label{fig:ym.eps}
\end{figure}

The gas physics in this simulation has been calibrated off of X-ray observations of cluster gas fractions as a function of temperature as discussed in \S\ref{sec:SZ} and \citet{BodeOV2009}.  For each halo with $\Mfof > 2\times10^{13}\ \Msunh$ and $z<3$ (whose gas model is described in \S \ref{sec:massivehalo}) we calculate $Y_{200}$ and $Y_{500}$, which are the projected Compton $y$ parameter in a disk of radius $R_{200}$ and $R_{500}$ respectively.  Here $R_{200}$ and $R_{500}$ are the radii within which the cluster mean mass density is 200 and 500 times the critical density at the cluster redshift.  We then rank the clusters by mass and bin them in mass bins of $\Delta \rm{log} (M_{200}) = 0.05$ and $\Delta \rm{log} (M_{500}) = 0.1$ for $M_{200}$ and $M_{500}$ respectively.   The mean of $M_{\Delta}$ as well as the mean and standard deviation on the mean of $Y_{\Delta} \times E(z)^{-2/3}$ are then calculated for each bin.  These values, for $M_{200}$ and $Y_{200}$, are plotted in Fig.~\ref{fig:ym.eps}, along with all the individual halos with $M_{200} > 2 \times 10^{14} M_{\odot}$ in the octant.

The binned values are then fit to the power-law relation
\begin{eqnarray}
\frac{Y_{\Delta}}{E(z)^{2/3}} = 10^{\beta} \Big( \frac{M_{\Delta}}{10^{14}
M_{\odot}} \Big)^{\alpha}. 
\label{eq:y_m}
\end{eqnarray}
For $M_{200} > 2 \times 10^{14} M_{\odot}$, we find for the $Y_{200}$ versus $M_{200}$ relation $\alpha =1.682 \pm 0.004$ and $\beta = -5.648 \pm 0.002$, with a reduced $\chi^{2}$ of 1.2 for $\sim 13,000$  clusters divided into 19 bins.  This slope is slightly steeper than that expected for the self-similar case.  It is also consistent with the $Y-M$ relations derived from hydrodynamical cluster simulations (e.g. \citealt{White2002,daSilva2004,Motl2005,Nagai2006,Shaw2008}).  The $Y-M$ relation calculated within a disk of radius $R_{2500}$ is also consistent with the observed $Y_{2500}-M_{\rm{gas}}(<R_{2500})$ relation of \citet{Bonamente2008}, as shown in \citet{BodeOV2009}.

\begin{center}
\begin{deluxetable*}{ccccccccc}
\tabletypesize{\scriptsize}
\tablecaption{Best- Fit Parameters for the $Y_{\Delta} - M_{\Delta}$ Relation, with $M_{\Delta} > 2 \times 10^{14} M_{\odot}$,
Fitted Using the Power Law Given in Equation \ref{eq:y_m} \label{tab:ym}}
\tablehead{\colhead{Relation} & \colhead{$\alpha$} & \colhead{$\sigma_{\alpha}$} & \colhead{$\beta$} & \colhead{$\sigma_{\beta}$} & \colhead{Reduced $\chi^{2}$} & \colhead{Clusters} & \colhead{Bins} & \colhead{$\sigma_{YM}$} 
}
\startdata
$Y_{200}-M_{200}$ & 1.682 & 0.004 & -5.648 & 0.002 & 1.2 & 12970 & 19 & 0.14\\
$Y_{500}-M_{500}$ & 1.668 & 0.009 & -5.509 & 0.004 & 1.9 & 4900 & 8 & 0.17 \\
$Y_{500}-M_{200}$ & 1.693 & 0.004 & -5.765 & 0.002 & 1.8 & 12970 & 19  & 0.13
\enddata
\end{deluxetable*}
\end{center}

We also find the scatter in this $Y-M$ relation about the best-fit relation.  Specifically we calculate
\begin{eqnarray}
\sigma_{YM} = \left [ \frac{\sum^{N}_{i=1} (\rm{ln} \tilde{Y}_{\Delta} - \rm{ln} Y_{\Delta})_i^2 }{N-2} \right ]^{1/2}
\end{eqnarray}
as discussed in e.g. \citet{Shaw2008}, where $ \tilde{Y}_{\Delta}$ is the projected SZ flux measured within a disk of radius $R_{\Delta}$, $Y_{\Delta}$ is the fitted SZ flux of for a cluster of mass $M_{\Delta}$ using the best-fit parameters for Eq. \ref{eq:y_m}, and N is the total number of clusters.  We find a scatter for the $Y_{200}-M_{200}$ relation of about $14\%$, and a slightly larger scatter of about $17\%$ for the $Y_{500}-M_{500}$ relation.  We find a lower scatter of roughly $13\%$ for the $Y_{500}-M_{200}$ relation, which has been suggested by \citet{Shaw2008} to have less scatter than the $Y_{200}-M_{200}$ relation.  We list the best-fit parameters for Eq. \ref{eq:y_m} and the scatter for the $Y_{200}-M_{200}$, $Y_{500}-M_{500}$, and $Y_{500}-M_{200}$ relations in Table~\ref{tab:ym}.  

Concerning the scatter given above, we assume hydrostatic equilibrium for the gas;  however, we do not assume the same entropy level for all the halos. The kinetic
energy of the gas is derived from that of the dark matter, so for a non-virialized, merging halo with a high kinetic energy relative to the cluster potential, the gas   
will also be given a high energy and thus temperature.  Thus merging of halos is taken into account at some level, although a full treatment of the gas physics would  
give more scatter.  Also, the repetition of halos as the periodic box is tiled to fill the octant, even though these halos are often at different redshifts and evolutionary states, may serve to slightly decrease the scatter.

Note that the $Y$ values discussed above are intrinsic to the cluster and do not include any contamination from foreground or background SZ signals along the line of sight.  Inclusion of this projection contamination will increase the $Y-M$ scatter somewhat, but for clusters with masses greater than roughly $2 \times 10^{14} M_{\odot}$, an optimistic mass limit of current SZ surveys, this projection contamination should be below $\sim 20\%$ over a wide range of possible $\sigma_8$ values (0.7-1.0) and redshifts ($z <2$) \citep{Holder2007,Hallman2007}.

\subsection{Point Source Contamination of SZ Clusters}

\begin{figure}[ht]
\epsscale{1.2}
\plotone{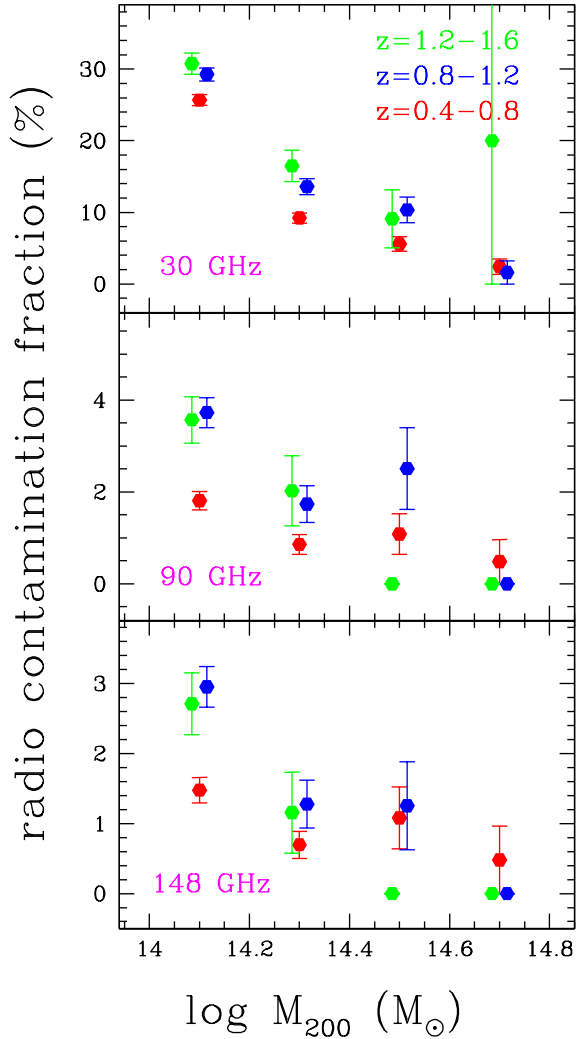}
\caption{Fraction of halos in this simulation that host radio galaxies whose combined flux equals $20\%$ or more of the halo's SZ flux decrement at 30 GHz (top), 90 GHz (middle), and 148 GHz (bottom).  Bin widths are $\Delta  \rm{log} (M_{200}) = 0.2$, and the error bars assume a Poisson distribution.  Red, blue, and green dots indicate the redshift ranges of $0.4-0.8$, $0.8-1.2$, and $1.2-1.6$ respectively. }\label{fig:RadioContam.eps}
\end{figure}

An important question for both determinations of an SZ cluster sample's selection function and measurements of the SZ power spectrum is the 
correlation and contamination of the SZ cluster signal by infrared and radio galaxies.  
Observations suggest we should expect some preference for radio and infrared galaxies to reside in galaxy clusters (e.g. \citealt{Coble2007,Lin2007, Daddi2009}), and these sources could potentially fill in the SZ decrement at frequencies below the null.  This point source contamination of the SZ flux could both affect cluster detection and bias a measured SZ flux - mass scaling relation.
If these effects are significant and unaccounted for, then inaccurate cosmological conclusions would result from measured SZ cluster samples.  Regarding measurements of the SZ power spectrum and derived values of $\sigma_8$, typically a significant number of high flux point sources are masked out of microwave maps before determining the power spectrum.  If these sources reside preferentially in massive clusters, a significant portion of the SZ signal could be masked out and a biased determination of $\sigma_8$ would arise.   

\begin{figure}[b]
\epsscale{1.2}
\plotone{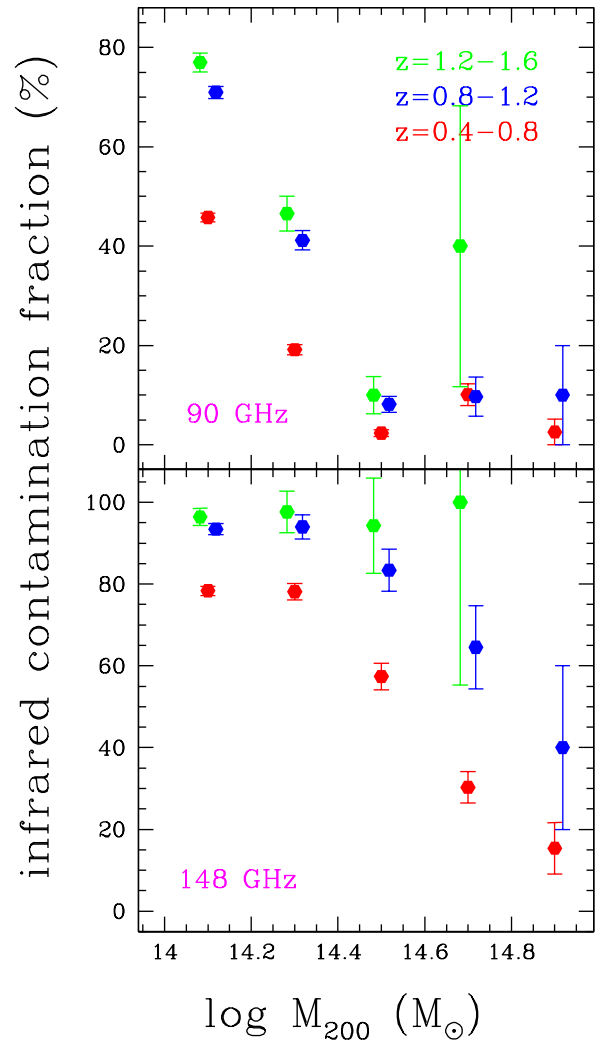}
\caption{Fraction of halos in this simulation that host infrared galaxies whose combined flux equals $20\%$ or more of the halo's SZ flux decrement at 90 GHz (top) and 148 GHz (bottom).  Bin widths and redshift ranges are the same as in Fig.~\ref{fig:RadioContam.eps}. An absence of an indicated fraction indicates no halos in that mass and redshift interval. }\label{fig:IRContam.eps}
\end{figure}

\begin{figure}[t]
\epsscale{1.2}
\plotone{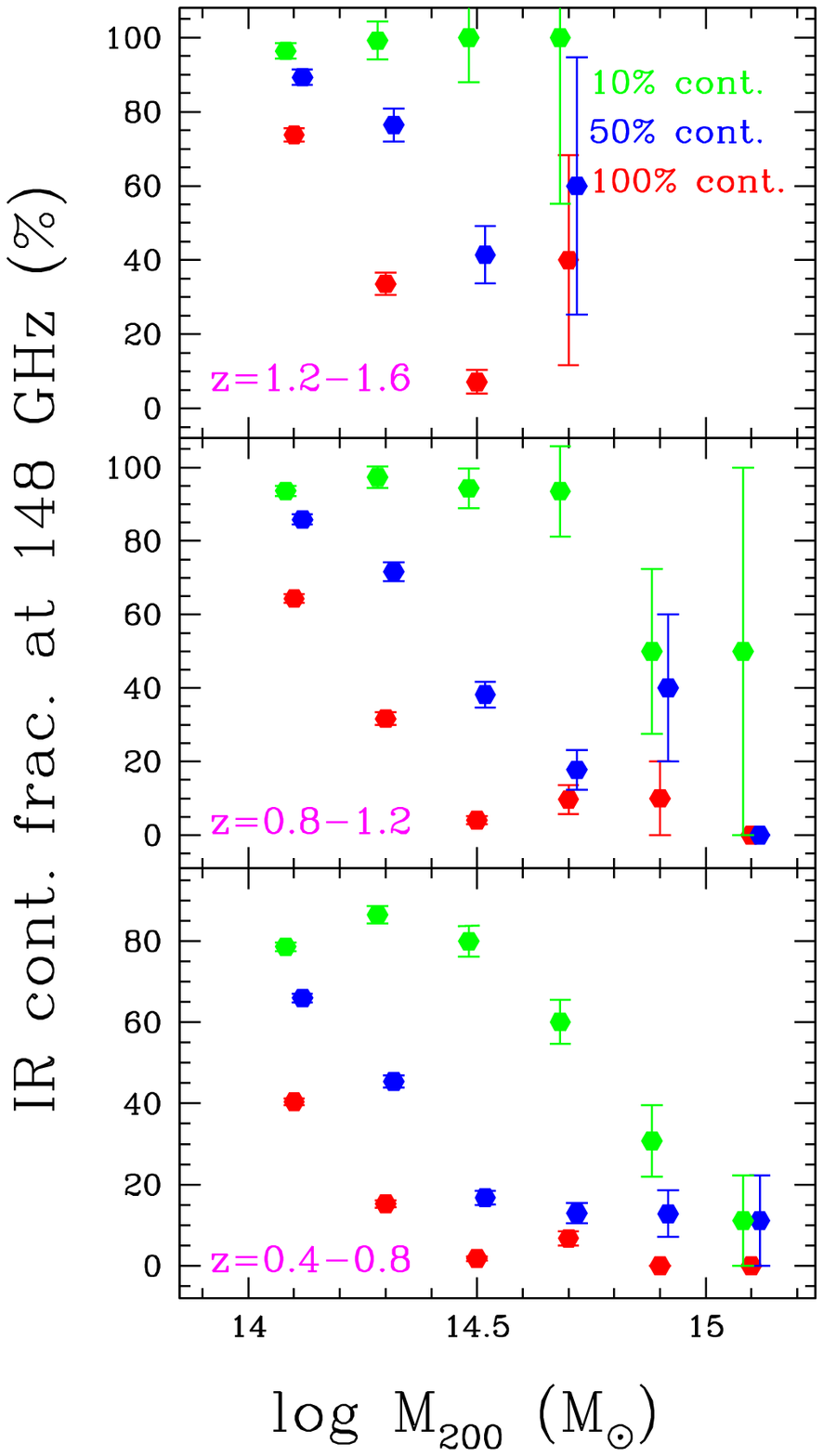}
\caption{Fraction of halos in this simulation that host infrared galaxies whose combined flux is greater than $10\%$ (green), $50\%$ (blue), or $100\%$ (red) of the halo's SZ flux decrement at 148 GHz.  Bin widths are the same as in Fig.~\ref{fig:RadioContam.eps}.  This is shown for the redshift ranges of $1.2-1.6$ (top), $0.8-1.2$ (middle), and $0.4-0.8$ (bottom). }\label{fig:IRContam148.eps}
\end{figure}

Here we investigate, via these simulations, the fraction of SZ clusters that may be substantially contaminated by radio or infrared galaxy populations.  We first add the flux of all galaxies of a given population residing in a given halo.  Then, using each galaxy cluster's integrated SZ flux within $R_{200}$ (in units of mJy), we calculate the fraction of halos whose combined flux from resident galaxies equals $20\%$ or more of the halo's SZ flux decrement at 30, 90, and 148 GHz.  Such halos will be harder to detect by SZ surveys, and their signal is more likely to be masked out in attempts to measure the SZ power spectrum.  We choose a $20\%$ contamination fraction as that is roughly where the uncertainty in the expected SZ flux for a cluster of a given mass becomes dominated by point source contamination as opposed to the intrinsic scatter in the $Y-M$ relation or projection effects.

For the contamination due to radio sources, we find that for halos with $M_{200} > 1 \times 10^{14} M_{\odot}$, roughly $3\%$ or less of clusters with $z<1.6$ have their SZ flux decrement at 148 GHz filled in by more than $20\%$ (see Fig.~\ref{fig:RadioContam.eps}).  At 90 GHz, this fraction increases slightly, and less than $4\%$ of clusters with $z<1.6$ have a contamination of $20\%$ or more.  At 30 GHz, one may expect 20 to $30\%$ of clusters with $M_{200} \sim 1 \times 10^{14} M_{\odot}$ to be so contaminated, with a decrease in this fraction as the halo mass is increased.  These contamination fractions exhibit a weak redshift dependence, with slightly larger fractions at higher redshifts.

Infrared galaxies seem to be a significant source of contamination for SZ clusters at 90 and 148 GHz (see Figs.~\ref{fig:IRContam.eps} and \ref{fig:IRContam148.eps}).  However, these simulations suggest that, at 90 GHz, the fraction of halos with $M_{200} > 2.5 \times 10^{14} M_{\odot}$ and $z<1.2$, that are contaminated at a level of $20\%$ or more, is less than $20\%$.  In all mass bins, the general trend is for the contamination fraction to increase at higher redshifts where more infrared galaxies reside.  There is no appreciable contamination from infrared galaxies at 30 GHz.  At 148 GHz, less than $20\%$ of halos with $M_{200} > 2.5 \times 10^{14} M_{\odot}$ and $z<0.8$ have their SZ decrements filled in at a level of $50\%$ or more (see Fig.~\ref{fig:IRContam148.eps}).  It should be noted that the above estimates only include galaxies that reside in halos and does not include contamination from galaxies along a halo's line of sight.  In this regard, these fractions should be viewed as lower limits.  The one exception to this is for the very high flux infrared galaxies that most likely represent a lensed population of galaxies.  These galaxies are placed in massive halos and are the main source of contamination in the most massive bins.  These sources should more properly be placed behind massive halos, however, the contamination effect in projection will still be similar.  

Such contamination fractions as discussed above should to be factored into sample completeness estimates and the scatter in the $Y-M$ relation in order to use SZ detected clusters for precision cosmology.  The models presented here should give a reasonable approximation to the expected levels of contamination from the different galaxy populations.  However, these estimates should be cross-checked and made more precise with multi-wavelength observations of SZ clusters.

\section{CONCLUSION} \label{sec:concl}

The simulations discussed here have been created for the use of the ACT team to test their data analysis pipeline.  To this end, we have modeled the primary astrophysical signals as realistically as possible by matching closely to recent observations.  We have also expanded the frequency coverage beyond the ACT bands and created maps and catalogs in an easily accessible format.  As such, other current and upcoming microwave background experiments may find these simulations useful for a variety of purposes. 

From the gas prescription in these simulations, which has been matched to recent X-ray observations, we find a scaling relation between Compton $y$ parameter and mass that is only slightly steeper than self-similar and has a scatter of $14\%$ to $17\%$ around the best-fit relation for $Y_{200}-M_{200}$ and $Y_{500}-M_{500}$ respectively.  This low intrinsic scatter suggests that measurement of the cluster SZ flux may be a relatively clean cluster mass proxy.  The correlation of radio galaxies with SZ clusters suggests that at 148 GHz (90 GHz), for clusters with $M_{200} > 10^{14} M_{\odot}$, less than $3\%$ ($4\%$) of these clusters will have their SZ decrement filled in by $20\%$ or more.  We find the contamination levels higher for infrared galaxies.  However, at 90 GHz, less than 20$\%$ of clusters with $M_{200} > 2.5 \times 10^{14} M_{\odot}$ and $z<1.2$ have their SZ decrements filled in at a level of $20\%$ or more. At 148 GHz, less than 20$\%$ of clusters with $M_{200} > 2.5 \times 10^{14} M_{\odot}$ and $z<0.8$ have their SZ decrements filled in at a level of 50$\%$ or larger.  We also find that a population of very high flux infrared galaxies, which may have been lensed, contribute most to the SZ contamination of very massive clusters at 90 and 148 GHz.  Multi-wavelength observations of SZ clusters will help to make these contamination estimates more precise.  This information is crucial for determining SZ cluster selection functions and the actual scatter in the observed $Y-M$ relation, which in turn are vital to efforts to extract cosmology from SZ cluster counts.  The inclusion of the lensing of the microwave background in a manner internally consistent with the large and small-scale structure is necessary to verify the accurate recovery of cosmology from cross-correlation studies between the lensed microwave background and tracers of large-scale structure.  Moreover, the simulations as a whole, matched to the most recent observations, will inform, through a variety of means, measurement of the angular power spectrum, higher point correlation functions, and the cosmology thus derived.  

Cosmic microwave background experiments are entering an exciting time of finer resolution and higher sensitivity which will open up the Sunyaev-Zel'dovich Universe, microwave background lensing, an assortment of cross-correlation studies, and deeper probes of inflation through tighter parameter constraints.  These simulations should provide a useful tool to aid analyses in these new and promising areas of discovery.

\acknowledgments
NS thanks Steve Allen, Olivier Dore, Joanna Dunkley, Joe Fowler, Gil Holder, Andrew Lawrence, Toby Marriage, 
Danica Marsden, Lyman Page, and David Spergel for useful discussions. 
SD thanks David Spergel, Chris Hirata, Paul Bode and Hy Trac for help and guidance during the development of the original lensing code.
CHM is grateful to R.E. Smith for insightful comments on the power spectrum modeling.
HT thanks Kavi Moodley and Ryan Warne for helpful discussions. 
NS is supported by the U.S. Department of Energy contract to SLAC no. DE-AC3-76SF00515.
PB was partially supported by NSF grant 0707731.
SD acknowledges  support from NASA ATP Grant NNX08AH30G.
YTL acknowledges support from the World Premier International Research Center Initiative, MEXT, Japan.
HT is supported by the Institute for Theory and Computation Fellowship.
Computer simulations and analysis were supported by the
National Science Foundation through TeraGrid resources
provided by the Pittsburgh Supercomputing Center
and the National Center for Supercomputing Applications
under grant AST070015; computations were also
performed at the TIGRESS high performance
computer center at Princeton University, which is jointly supported by
the Princeton Institute for Computational Science and Engineering and
the Princeton University Office of Information Technology.  
We acknowledge support from the PIRE program and NSF Grant OISE/0530095.
Some of the results in this paper have been derived using the HEALPix package \citep{Gorski2005}.  
We also acknowledge the use of the Legacy Archive for 
Microwave Background Data Analysis (LAMBDA). Support for LAMBDA is provided by the NASA Office of Space Science.

\bibliographystyle{apj}

\begin{thebibliography}{}

\bibitem[\protect\citeauthoryear{{Acquaviva} et~al.}{{Acquaviva}
  et~al.}{2008}]{acquaviva.hajian.ea:2008}
{Acquaviva}, V., {Hajian}, A., {Spergel}, D.~N.,  \& {Das}, S. 2008, \prd, 78,
  043514

\bibitem[\protect\citeauthoryear{{Amblard}, {Vale}, \& {White}}{{Amblard}
  et~al.}{2004}]{amblard.vale.ea:2004}
{Amblard}, A., {Vale}, C.,  \& {White}, M. 2004, New Astronomy, 9, 687

\bibitem[\protect\citeauthoryear{{Austermann} et~al.}{{Austermann}
  et~al.}{2009}]{Austermann2009}
{Austermann}, J.~E., et~al. 2009, ArXiv e-prints

\bibitem[\protect\citeauthoryear{{Basak}, {Prunet}, \& {Benabed}}{{Basak}
  et~al.}{2008}]{basak.prunet.ea:2008}
{Basak}, S., {Prunet}, S.,  \& {Benabed}, K. 2008, ArXiv e-prints

\bibitem[\protect\citeauthoryear{{Bennett} et~al.}{{Bennett}
  et~al.}{2003}]{Bennett2003b}
{Bennett}, C.~L., et~al. 2003, \apjs, 148, 97

\bibitem[\protect\citeauthoryear{{Blain}}{{Blain}}{1999}]{blain99}
{Blain}, A.~W. 1999, \mnras, 309, 955

\bibitem[\protect\citeauthoryear{{Blundell}, {Rawlings}, \&
  {Willott}}{{Blundell} et~al.}{1999}]{blundell99}
{Blundell}, K.~M., {Rawlings}, S.,  \& {Willott}, C.~J. 1999, \aj, 117, 677

\bibitem[\protect\citeauthoryear{{Bode} \& {Ostriker}}{{Bode} \&
  {Ostriker}}{2003}]{bode.ostriker:2003}
{Bode}, P.,  \& {Ostriker}, J.~P. 2003, \apjs, 145, 1

\bibitem[\protect\citeauthoryear{{Bode}, {Ostriker}, \& {Vikhlinin}}{{Bode}
  et~al.}{2009}]{BodeOV2009}
{Bode}, P., {Ostriker}, J.~P.,  \& {Vikhlinin}, A. 2009, ArXiv e-prints

\bibitem[\protect\citeauthoryear{{Bode} et~al.}{{Bode}
  et~al.}{2007}]{bode.ostriker.ea:2007}
{Bode}, P., {Ostriker}, J.~P., {Weller}, J.,  \& {Shaw}, L. 2007, \apj, 663,
  139

\bibitem[\protect\citeauthoryear{{Bode}, {Ostriker}, \& {Xu}}{{Bode}
  et~al.}{2000}]{bode.ostriker.ea:2000}
{Bode}, P., {Ostriker}, J.~P.,  \& {Xu}, G. 2000, \apjs, 128, 561

\bibitem[\protect\citeauthoryear{{Bonamente} et~al.}{{Bonamente}
  et~al.}{2008}]{Bonamente2008}
{Bonamente}, M., {Joy}, M., {LaRoque}, S.~J., {Carlstrom}, J.~E., {Nagai}, D.,
  \& {Marrone}, D.~P. 2008, \apj, 675, 106

\bibitem[\protect\citeauthoryear{{Bondi} et~al.}{{Bondi}
  et~al.}{2003}]{bondi03}
{Bondi}, M., et~al. 2003, \aap, 403, 857

\bibitem[\protect\citeauthoryear{{Carbone} et~al.}{{Carbone}
  et~al.}{2008}]{carbone.baccigalupi.ea:2008}
{Carbone}, C., {Baccigalupi}, C., {Bartelmann}, M., {Matarrese}, S.,  \&
  {Springel}, V. 2008, ArXiv e-prints

\bibitem[\protect\citeauthoryear{{Carbone} et~al.}{{Carbone}
  et~al.}{2009}]{Carbone2009}
{Carbone}, C., {Baccigalupi}, C., {Bartelmann}, M., {Matarrese}, S.,  \&
  {Springel}, V. 2009, \mnras, 576

\bibitem[\protect\citeauthoryear{{Carbone} et~al.}{{Carbone}
  et~al.}{2007}]{carbone.springel.ea:2007}
{Carbone}, C., {Springel}, V., {Baccigalupi}, C., {Bartelmann}, M.,  \&
  {Matarrese}, S. 2007, ArXiv e-prints

\bibitem[\protect\citeauthoryear{{Cen} \& {Ostriker}}{{Cen} \&
  {Ostriker}}{1999}]{CenOstriker1999}
{Cen}, R.,  \& {Ostriker}, J.~P. 1999, \apj, 514, 1

\bibitem[\protect\citeauthoryear{{Cleary} et~al.}{{Cleary}
  et~al.}{2005}]{cleary05}
{Cleary}, K.~A., et~al. 2005, \mnras, 360, 340

\bibitem[\protect\citeauthoryear{{Coble} et~al.}{{Coble}
  et~al.}{2007}]{Coble2007}
{Coble}, K., et~al. 2007, \aj, 134, 897

\bibitem[\protect\citeauthoryear{{Cohn} \& {White}}{{Cohn} \&
  {White}}{2009}]{CohnWhite2009}
{Cohn}, J.~D.,  \& {White}, M. 2009, \mnras, 393, 393

\bibitem[\protect\citeauthoryear{{Coppin} et~al.}{{Coppin}
  et~al.}{2006}]{coppin}
{Coppin}, K., et~al. 2006, \mnras, 372, 1621

\bibitem[\protect\citeauthoryear{{da Silva} et~al.}{{da Silva}
  et~al.}{2004}]{daSilva2004}
{da Silva}, A.~C., {Kay}, S.~T., {Liddle}, A.~R.,  \& {Thomas}, P.~A. 2004,
  \mnras, 348, 1401

\bibitem[\protect\citeauthoryear{{Daddi} et~al.}{{Daddi}
  et~al.}{2009}]{Daddi2009}
{Daddi}, E., et~al. 2009, \apj, 694, 1517

\bibitem[\protect\citeauthoryear{{Das} \& {Bode}}{{Das} \&
  {Bode}}{2008}]{das.bode:2008}
{Das}, S.,  \& {Bode}, P. 2008, \apj, 682, 1

\bibitem[\protect\citeauthoryear{Das \& Spergel}{Das \&
  Spergel}{2009}]{das.spergel:2009*1}
Das, S.,  \& Spergel, D.~N. 2009, Physical Review D (Particles, Fields,
  Gravitation, and Cosmology), 79, 043509

\bibitem[\protect\citeauthoryear{{Dav{\'e}} et~al.}{{Dav{\'e}}
  et~al.}{2001}]{Dave2001}
{Dav{\'e}}, R., et~al. 2001, \apj, 552, 473

\bibitem[\protect\citeauthoryear{{de Putter}, {Zahn}, \& {Linder}}{{de Putter}
  et~al.}{2009}]{.zahn.ea:2009}
{de Putter}, R., {Zahn}, O.,  \& {Linder}, E.~V. 2009, ArXiv e-prints

\bibitem[\protect\citeauthoryear{{de Zotti} et~al.}{{de Zotti}
  et~al.}{2005}]{dezotti05}
{de Zotti}, G., {Ricci}, R., {Mesa}, D., {Silva}, L., {Mazzotta}, P.,
  {Toffolatti}, L.,  \& {Gonz{\' a}lez-Nuevo}, J. 2005, \aap, 431, 893

\bibitem[\protect\citeauthoryear{{Devlin} et~al.}{{Devlin}
  et~al.}{2009}]{Devlin2009}
{Devlin}, M.~J., et~al. 2009, \nat, 458, 737

\bibitem[\protect\citeauthoryear{{Diaferio} et~al.}{{Diaferio}
  et~al.}{2005}]{Diaferio2005}
{Diaferio}, A., et~al. 2005, \mnras, 356, 1477

\bibitem[\protect\citeauthoryear{{Dunkley} et~al.}{{Dunkley}
  et~al.}{2009}]{Dunkley2009}
{Dunkley}, J., et~al. 2009, \apjs, 180, 306

\bibitem[\protect\citeauthoryear{{Dunne} et~al.}{{Dunne} et~al.}{2000}]{dunne}
{Dunne}, L., {Eales}, S., {Edmunds}, M., {Ivison}, R., {Alexander}, P.,  \&
  {Clements}, D.~L. 2000, \mnras, 315, 115

\bibitem[\protect\citeauthoryear{{Dye} et~al.}{{Dye}
  et~al.}{2009}]{blast_radio}
{Dye}, S., et~al. 2009, ArXiv e-prints

\bibitem[\protect\citeauthoryear{{Fanaroff} \& {Riley}}{{Fanaroff} \&
  {Riley}}{1974}]{fanaroff74}
{Fanaroff}, B.~L.,  \& {Riley}, J.~M. 1974, \mnras, 167, 31P

\bibitem[\protect\citeauthoryear{{Finkbeiner}, {Davis}, \&
  {Schlegel}}{{Finkbeiner} et~al.}{1999}]{Finkbeiner1999}
{Finkbeiner}, D.~P., {Davis}, M.,  \& {Schlegel}, D.~J. 1999, \apj, 524, 867

\bibitem[\protect\citeauthoryear{{Fixsen} et~al.}{{Fixsen}
  et~al.}{1998}]{firas}
{Fixsen}, D.~J., {Dwek}, E., {Mather}, J.~C., {Bennett}, C.~L.,  \& {Shafer},
  R.~A. 1998, \apj, 508, 123

\bibitem[\protect\citeauthoryear{{Gawiser} et~al.}{{Gawiser}
  et~al.}{1998}]{Gawiser1998}
{Gawiser}, E., et~al. 1998, ArXiv Astrophysics e-prints

\bibitem[\protect\citeauthoryear{{Gawiser} \& {Smoot}}{{Gawiser} \&
  {Smoot}}{1997}]{Gawiser1997}
{Gawiser}, E.,  \& {Smoot}, G.~F. 1997, \apjl, 480, L1

\bibitem[\protect\citeauthoryear{{G{\'o}rski} et~al.}{{G{\'o}rski}
  et~al.}{2005}]{Gorski2005}
{G{\'o}rski}, K.~M., {Hivon}, E., {Banday}, A.~J., {Wandelt}, B.~D., {Hansen},
  F.~K., {Reinecke}, M.,  \& {Bartelmann}, M. 2005, \apj, 622, 759

\bibitem[\protect\citeauthoryear{{Haiman} \& {Knox}}{{Haiman} \&
  {Knox}}{2000}]{zoltan00}
{Haiman}, Z.,  \& {Knox}, L. 2000, \apj, 530, 124

\bibitem[\protect\citeauthoryear{{Hallman} et~al.}{{Hallman}
  et~al.}{2007}]{Hallman2007}
{Hallman}, E.~J., {O'Shea}, B.~W., {Burns}, J.~O., {Norman}, M.~L., {Harkness},
  R.,  \& {Wagner}, R. 2007, \apj, 671, 27

\bibitem[\protect\citeauthoryear{{Hallman} et~al.}{{Hallman}
  et~al.}{2009}]{Hallman2009}
{Hallman}, E.~J., {O'Shea}, B.~W., {Smith}, B.~D., {Burns}, J.~O.,  \&
  {Norman}, M.~L. 2009, \apj, 698, 1795

\bibitem[\protect\citeauthoryear{{Hern{\'a}ndez-Monteagudo}
  et~al.}{{Hern{\'a}ndez-Monteagudo} et~al.}{2006}]{CHM2006}
{Hern{\'a}ndez-Monteagudo}, C., {Trac}, H., {Verde}, L.,  \& {Jimenez}, R.
  2006, \apjl, 652, L1

\bibitem[\protect\citeauthoryear{{Hinshaw} et~al.}{{Hinshaw}
  et~al.}{2009}]{Hinshaw2009}
{Hinshaw}, G., et~al. 2009, \apjs, 180, 225

\bibitem[\protect\citeauthoryear{{Hirata} \& {Seljak}}{{Hirata} \&
  {Seljak}}{2003}]{hirata.seljak:2003*1}
{Hirata}, C.~M.,  \& {Seljak}, U. 2003, \prd, 67, 043001

\bibitem[\protect\citeauthoryear{{Holder}, {McCarthy}, \& {Babul}}{{Holder}
  et~al.}{2007}]{Holder2007}
{Holder}, G.~P., {McCarthy}, I.~G.,  \& {Babul}, A. 2007, \mnras, 382, 1697

\bibitem[\protect\citeauthoryear{{Hopkins} \& {Beacom}}{{Hopkins} \&
  {Beacom}}{2006}]{hopkinsb}
{Hopkins}, A.~M.,  \& {Beacom}, J.~F. 2006, \apj, 651, 142

\bibitem[\protect\citeauthoryear{{Hu} \& {Okamoto}}{{Hu} \&
  {Okamoto}}{2002}]{hu.okamoto:2002}
{Hu}, W.,  \& {Okamoto}, T. 2002, \apj, 574, 566

\bibitem[\protect\citeauthoryear{{Huynh} et~al.}{{Huynh}
  et~al.}{2005}]{huynh05}
{Huynh}, M.~T., {Jackson}, C.~A., {Norris}, R.~P.,  \& {Prandoni}, I. 2005,
  \aj, 130, 1373

\bibitem[\protect\citeauthoryear{{Jackson} \& {Wall}}{{Jackson} \&
  {Wall}}{1999}]{jackson99}
{Jackson}, C.~A.,  \& {Wall}, J.~V. 1999, \mnras, 304, 160

\bibitem[\protect\citeauthoryear{{Jenkins} et~al.}{{Jenkins}
  et~al.}{2001}]{Jenkins2001}
{Jenkins}, A., {Frenk}, C.~S., {White}, S.~D.~M., {Colberg}, J.~M., {Cole}, S.,
  {Evrard}, A.~E., {Couchman}, H.~M.~P.,  \& {Yoshida}, N. 2001, \mnras, 321,
  372

\bibitem[\protect\citeauthoryear{{Katgert}, {Oort}, \& {Windhorst}}{{Katgert}
  et~al.}{1988}]{katgert88}
{Katgert}, P., {Oort}, M.~J.~A.,  \& {Windhorst}, R.~A. 1988, \aap, 195, 21

\bibitem[\protect\citeauthoryear{{Kennicutt}}{{Kennicutt}}{1998}]{kenni}
{Kennicutt}, R.~C., Jr. 1998, \araa, 36, 189

\bibitem[\protect\citeauthoryear{{Knox}, {Holder}, \& {Church}}{{Knox}
  et~al.}{2004}]{knox04}
{Knox}, L., {Holder}, G.~P.,  \& {Church}, S.~E. 2004, \apj, 612, 96

\bibitem[\protect\citeauthoryear{{Komatsu} et~al.}{{Komatsu}
  et~al.}{2009}]{Komatsu2009}
{Komatsu}, E., et~al. 2009, \apjs, 180, 330

\bibitem[\protect\citeauthoryear{{Komatsu} \& {Seljak}}{{Komatsu} \&
  {Seljak}}{2002}]{Komatsu2002}
{Komatsu}, E.,  \& {Seljak}, U. 2002, \mnras, 336, 1256

\bibitem[\protect\citeauthoryear{{Lagache} et~al.}{{Lagache}
  et~al.}{2007}]{Lagache2007}
{Lagache}, G., {Bavouzet}, N., {Fernandez-Conde}, N., {Ponthieu}, N., {Rodet},
  T., {Dole}, H., {Miville-Desch{\^e}nes}, M.-A.,  \& {Puget}, J.-L. 2007,
  \apjl, 665, L89

\bibitem[\protect\citeauthoryear{{Laing}, {Riley}, \& {Longair}}{{Laing}
  et~al.}{1983}]{laing83}
{Laing}, R.~A., {Riley}, J.~M.,  \& {Longair}, M.~S. 1983, \mnras, 204, 151

\bibitem[\protect\citeauthoryear{{Leach} et~al.}{{Leach}
  et~al.}{2008}]{Leach2008}
{Leach}, S.~M., et~al. 2008, \aap, 491, 597

\bibitem[\protect\citeauthoryear{{Lewis}}{{Lewis}}{2005}]{lewis:2005}
{Lewis}, A. 2005, \prd, 71, 083008

\bibitem[\protect\citeauthoryear{{Lewis} \& {Challinor}}{{Lewis} \&
  {Challinor}}{2006}]{lewis.challinor:2006}
{Lewis}, A.,  \& {Challinor}, A. 2006, \physrep, 429, 1

\bibitem[\protect\citeauthoryear{{Lima}, {Jain}, \& {Devlin}}{{Lima}
  et~al.}{2009}]{Lima2009}
{Lima}, M., {Jain}, B.,  \& {Devlin}, M. 2009, ArXiv e-prints

\bibitem[\protect\citeauthoryear{{Lin} et~al.}{{Lin} et~al.}{2009a}]{Lin2009}
{Lin}, Y.-T., et~al. 2009a, in prep

\bibitem[\protect\citeauthoryear{{Lin} \& {Mohr}}{{Lin} \&
  {Mohr}}{2007}]{Lin2007}
{Lin}, Y.-T.,  \& {Mohr}, J.~J. 2007, \apjs, 170, 71

\bibitem[\protect\citeauthoryear{{Lin} et~al.}{{Lin} et~al.}{2009b}]{lin09}
{Lin}, Y.-T., {Partridge}, B., {Pober}, J.~C., {Bouchefry}, K.~E., {Burke}, S.,
  {Klein}, J.~N., {Coish}, J.~W.,  \& {Huffenberger}, K.~M. 2009b, \apj, 694,
  992

\bibitem[\protect\citeauthoryear{{Malte Sch{\"a}fer} \& {Bartelmann}}{{Malte
  Sch{\"a}fer} \& {Bartelmann}}{2007}]{Schafer2007}
{Malte Sch{\"a}fer}, B.,  \& {Bartelmann}, M. 2007, \mnras, 377, 253

\bibitem[\protect\citeauthoryear{{Marsden} et~al.}{{Marsden}
  et~al.}{2009}]{Marsden2009}
{Marsden}, G., et~al. 2009, ArXiv e-prints

\bibitem[\protect\citeauthoryear{{McQuinn} et~al.}{{McQuinn}
  et~al.}{2005}]{McQuinn2005}
{McQuinn}, M., {Furlanetto}, S.~R., {Hernquist}, L., {Zahn}, O.,  \&
  {Zaldarriaga}, M. 2005, \apj, 630, 643

\bibitem[\protect\citeauthoryear{{Metzler}}{{Metzler}}{1998}]{Metzler1998}
{Metzler}, C.~A. 1998, ArXiv Astrophysics e-prints

\bibitem[\protect\citeauthoryear{{Motl} et~al.}{{Motl} et~al.}{2005}]{Motl2005}
{Motl}, P.~M., {Hallman}, E.~J., {Burns}, J.~O.,  \& {Norman}, M.~L. 2005,
  \apjl, 623, L63

\bibitem[\protect\citeauthoryear{{Nagai}}{{Nagai}}{2006}]{Nagai2006}
{Nagai}, D. 2006, \apj, 650, 538

\bibitem[\protect\citeauthoryear{{Nozawa}, {Itoh}, \& {Kohyama}}{{Nozawa}
  et~al.}{1998}]{Nozawa1998}
{Nozawa}, S., {Itoh}, N.,  \& {Kohyama}, Y. 1998, \apj, 508, 17

\bibitem[\protect\citeauthoryear{{Okamoto} \& {Hu}}{{Okamoto} \&
  {Hu}}{2003}]{okamoto.hu:2003}
{Okamoto}, T.,  \& {Hu}, W. 2003, \prd, 67, 083002

\bibitem[\protect\citeauthoryear{{Ostriker}, {Bode}, \& {Babul}}{{Ostriker}
  et~al.}{2005}]{OstrikerBB2005}
{Ostriker}, J.~P., {Bode}, P.,  \& {Babul}, A. 2005, \apj, 634, 964

\bibitem[\protect\citeauthoryear{{Ostriker} \& {Vishniac}}{{Ostriker} \&
  {Vishniac}}{1986}]{OstrikerVishniac1986}
{Ostriker}, J.~P.,  \& {Vishniac}, E.~T. 1986, \apjl, 306, L51

\bibitem[\protect\citeauthoryear{{Pace} et~al.}{{Pace} et~al.}{2008}]{Pace2008}
{Pace}, F., {Maturi}, M., {Bartelmann}, M., {Cappelluti}, N., {Dolag}, K.,
  {Meneghetti}, M.,  \& {Moscardini}, L. 2008, \aap, 483, 389

\bibitem[\protect\citeauthoryear{{Patanchon} et~al.}{{Patanchon}
  et~al.}{2009}]{blast_counts}
{Patanchon}, G., et~al. 2009, ArXiv e-prints

\bibitem[\protect\citeauthoryear{{Peel}, {Battye}, \& {Kay}}{{Peel}
  et~al.}{2009}]{Peel2009}
{Peel}, M.~W., {Battye}, R.~A.,  \& {Kay}, S.~T. 2009, ArXiv e-prints

\bibitem[\protect\citeauthoryear{{Perotto} et~al.}{{Perotto}
  et~al.}{2009}]{perotto.bobin.ea:2009}
{Perotto}, L., {Bobin}, J., {Plaszczynski}, S., {Starck}, J.~.,  \& {Lavabre},
  A. 2009, ArXiv e-prints

\bibitem[\protect\citeauthoryear{{Pfrommer} et~al.}{{Pfrommer}
  et~al.}{2007}]{Pfrommer2007}
{Pfrommer}, C., {En{\ss}lin}, T.~A., {Springel}, V., {Jubelgas}, M.,  \&
  {Dolag}, K. 2007, \mnras, 378, 385

\bibitem[\protect\citeauthoryear{{Reichardt} et~al.}{{Reichardt}
  et~al.}{2009}]{reichardt09}
{Reichardt}, C.~L., et~al. 2009, \apj, 694, 1200

\bibitem[\protect\citeauthoryear{{Righi}, {Hern{\'a}ndez-Monteagudo}, \&
  {Sunyaev}}{{Righi} et~al.}{2008}]{righi}
{Righi}, M., {Hern{\'a}ndez-Monteagudo}, C.,  \& {Sunyaev}, R.~A. 2008, \aap,
  478, 685

\bibitem[\protect\citeauthoryear{{Roncarelli} et~al.}{{Roncarelli}
  et~al.}{2007}]{Roncarelli2007}
{Roncarelli}, M., {Moscardini}, L., {Borgani}, S.,  \& {Dolag}, K. 2007,
  \mnras, 378, 1259

\bibitem[\protect\citeauthoryear{{Sadler} et~al.}{{Sadler}
  et~al.}{2008}]{sadler08}
{Sadler}, E.~M., {Ricci}, R., {Ekers}, R.~D., {Sault}, R.~J., {Jackson}, C.~A.,
   \& {de Zotti}, G. 2008, \mnras, 385, 1656

\bibitem[\protect\citeauthoryear{{Santos} et~al.}{{Santos}
  et~al.}{2003}]{Santos2003}
{Santos}, M.~G., {Cooray}, A., {Haiman}, Z., {Knox}, L.,  \& {Ma}, C. 2003,
  \apj, 598, 756

\bibitem[\protect\citeauthoryear{{Schlegel}, {Finkbeiner}, \&
  {Davis}}{{Schlegel} et~al.}{1998}]{Schlegel1998}
{Schlegel}, D.~J., {Finkbeiner}, D.~P.,  \& {Davis}, M. 1998, \apj, 500, 525

\bibitem[\protect\citeauthoryear{{Sehgal} et~al.}{{Sehgal}
  et~al.}{2007}]{Sehgal07}
{Sehgal}, N., {Trac}, H., {Huffenberger}, K.,  \& {Bode}, P. 2007, \apj, 664,
  149

\bibitem[\protect\citeauthoryear{{Shaw}, {Holder}, \& {Bode}}{{Shaw}
  et~al.}{2008}]{Shaw2008}
{Shaw}, L.~D., {Holder}, G.~P.,  \& {Bode}, P. 2008, \apj, 686, 206

\bibitem[\protect\citeauthoryear{{Sheth} \& {Tormen}}{{Sheth} \&
  {Tormen}}{1999}]{STbias}
{Sheth}, R.~K.,  \& {Tormen}, G. 1999, \mnras, 308, 119

\bibitem[\protect\citeauthoryear{{Smith} et~al.}{{Smith}
  et~al.}{2008}]{smith.cooray.ea:2008}
{Smith}, K.~M., et~al. 2008, ArXiv e-prints

\bibitem[\protect\citeauthoryear{{Smith} et~al.}{{Smith}
  et~al.}{2003}]{smith.peacock.ea:2003}
{Smith}, R.~E., et~al. 2003, \mnras, 341, 1311

\bibitem[\protect\citeauthoryear{{Smoot}}{{Smoot}}{1998}]{Smoot1998}
{Smoot}, G.~F. 1998, ArXiv Astrophysics e-prints

\bibitem[\protect\citeauthoryear{{Sokasian}, {Gawiser}, \& {Smoot}}{{Sokasian}
  et~al.}{2001}]{Sokasian2001}
{Sokasian}, A., {Gawiser}, E.,  \& {Smoot}, G.~F. 2001, \apj, 562, 88

\bibitem[\protect\citeauthoryear{{Song} et~al.}{{Song} et~al.}{2003}]{song03}
{Song}, Y.-S., {Cooray}, A., {Knox}, L.,  \& {Zaldarriaga}, M. 2003, \apj, 590,
  664

\bibitem[\protect\citeauthoryear{{Sun} et~al.}{{Sun} et~al.}{2009}]{Sun2009}
{Sun}, M., {Voit}, G.~M., {Donahue}, M., {Jones}, C., {Forman}, W.,  \&
  {Vikhlinin}, A. 2009, \apj, 693, 1142

\bibitem[\protect\citeauthoryear{{Sunyaev} \& {Zeldovich}}{{Sunyaev} \&
  {Zeldovich}}{1970}]{SZ1970}
{Sunyaev}, R.~A.,  \& {Zeldovich}, Y.~B. 1970, Comments on Astrophysics and
  Space Physics, 2, 66

\bibitem[\protect\citeauthoryear{{Sunyaev} \& {Zeldovich}}{{Sunyaev} \&
  {Zeldovich}}{1972}]{SZ1972}
{Sunyaev}, R.~A.,  \& {Zeldovich}, Y.~B. 1972, Comments on Astrophysics and
  Space Physics, 4, 173

\bibitem[\protect\citeauthoryear{{Toffolatti} et~al.}{{Toffolatti}
  et~al.}{1998}]{Toffolatti1998}
{Toffolatti}, L., {Argueso Gomez}, F., {de Zotti}, G., {Mazzei}, P.,
  {Franceschini}, A., {Danese}, L.,  \& {Burigana}, C. 1998, \mnras, 297, 117

\bibitem[\protect\citeauthoryear{{Trac}, {Bode}, \& {Ostriker}}{{Trac}
  et~al.}{2009}]{Trac09}
{Trac}, H., {Bode}, P.,  \& {Ostriker}, J.~P. 2009, in prep

\bibitem[\protect\citeauthoryear{{Vallinotto} et~al.}{{Vallinotto}
  et~al.}{2009}]{vallinotto.das.ea:2009}
{Vallinotto}, A., {Das}, S., {Spergel}, D.~N.,  \& {Viel}, M. 2009, ArXiv
  e-prints

\bibitem[\protect\citeauthoryear{{Verde}, {Heavens}, \& {Matarrese}}{{Verde}
  et~al.}{2000}]{verde00}
{Verde}, L., {Heavens}, A.~F.,  \& {Matarrese}, S. 2000, \mnras, 318, 584

\bibitem[\protect\citeauthoryear{{Viero} et~al.}{{Viero}
  et~al.}{2009}]{Viero09}
{Viero}, M.~P., et~al. 2009, ArXiv e-prints

\bibitem[\protect\citeauthoryear{{Vikhlinin} et~al.}{{Vikhlinin}
  et~al.}{2006}]{Vikhlinin06}
{Vikhlinin}, A., {Kravtsov}, A., {Forman}, W., {Jones}, C., {Markevitch}, M.,
  {Murray}, S.~S.,  \& {Van Speybroeck}, L. 2006, \apj, 640, 691

\bibitem[\protect\citeauthoryear{{Vishniac}}{{Vishniac}}{1987}]{Vishniac1987}
{Vishniac}, E.~T. 1987, \apj, 322, 597

\bibitem[\protect\citeauthoryear{{White}, {Hernquist}, \& {Springel}}{{White}
  et~al.}{2002}]{White2002}
{White}, M., {Hernquist}, L.,  \& {Springel}, V. 2002, \apj, 579, 16

\bibitem[\protect\citeauthoryear{{Wik} et~al.}{{Wik} et~al.}{2008}]{Wik2008}
{Wik}, D.~R., {Sarazin}, C.~L., {Ricker}, P.~M.,  \& {Randall}, S.~W. 2008,
  \apj, 680, 17

\bibitem[\protect\citeauthoryear{{Willott} et~al.}{{Willott}
  et~al.}{2001}]{willott01}
{Willott}, C.~J., {Rawlings}, S., {Blundell}, K.~M., {Lacy}, M.,  \& {Eales},
  S.~A. 2001, \mnras, 322, 536

\bibitem[\protect\citeauthoryear{{Wilman} et~al.}{{Wilman}
  et~al.}{2008}]{wilman08}
{Wilman}, R.~J., et~al. 2008, \mnras, 388, 1335

\bibitem[\protect\citeauthoryear{{Windhorst} et~al.}{{Windhorst}
  et~al.}{1993}]{windhorst93}
{Windhorst}, R.~A., {Fomalont}, E.~B., {Partridge}, R.~B.,  \& {Lowenthal},
  J.~D. 1993, \apj, 405, 498

\bibitem[\protect\citeauthoryear{{Wright} et~al.}{{Wright}
  et~al.}{2009}]{wright09}
{Wright}, E.~L., et~al. 2009, \apjs, 180, 283

\bibitem[\protect\citeauthoryear{{Yoo} \& {Zaldarriaga}}{{Yoo} \&
  {Zaldarriaga}}{2008}]{yoo.zaldarriaga:2008}
{Yoo}, J.,  \& {Zaldarriaga}, M. 2008, ArXiv e-prints, 805

\bibitem[\protect\citeauthoryear{{Zahn} et~al.}{{Zahn} et~al.}{2005}]{Zahn2005}
{Zahn}, O., {Zaldarriaga}, M., {Hernquist}, L.,  \& {McQuinn}, M. 2005, \apj,
  630, 657

\end{thebibliography}

\end{document}